# The Medieval Climate Anomaly, the Oort Minimum and socio-political dynamics in the Eastern Mediterranean and the Byzantine Empire, 10th to 12th century

Johannes Preiser-Kapeller

**Pre-Print, to be published in:** Adam Izdebski/Johannes Preiser-Kapeller (eds.), A Companion to the Environmental History of Byzantium (Brill's Companions to the Byzantine World). Leiden 2023 (under review)

The following chapter examines the palaeoclimatic background and the regional manifestations of the so-called "Medieval Climate Anomaly" in the Eastern Mediterranean, with a focus on the Byzantine Empire, but also including neighbouring polities. It explores the interplay between climatic factors and the socio-economic dynamics between the 10th and 12th centuries, concentrating on the late 10th and 11th centuries (also overlapping with the "Oort Solar Minimum"). In particular, it contrasts scenarios of an "economic boom" and of an "collapse of the Eastern Mediterranean" created in recent scholarship for this period and evaluates these notions based on a close reading (and citation) of historiographical and other written sources. Thereby, both potentials as well as problems of a combination of "archives of society" and "archives of nature" become evident.[1]

## 1 From the "Medieval Warm Period" to the "Medieval Climate Anomaly"

The intellectual history of what was called the "Medieval Warm Period" began in 1959, when Hubert Lamb (1913-1997), a pioneer of historical climatology, introduced the term for the time between the 10th and 13th centuries CE based on his reading of medieval sources and the then limited number of temperature reconstructions for England.[2] In 1964, using the same data, Lamb published an influential graph with the caption "temperatures (°C) prevailing in central England, 50 years averages", indicating that it represented rather rough estimates. Nevertheless, the suggestive power of the image, showing an impressive temperature mount of a "Medieval Warm Period" proved more powerful than the caveats of Lamb.[3]

---

[1] For these two types of "archives", see Pfister/Wanner, *Klima und Gesellschaft in Europa*, pp. 15–20 and 117–47.
[2] Rohr/Camenisch/Pribyl, "European Middle Ages"; Summerhayes, *Palaeoclimatology*, p. 440; Pfister/Wanner, *Klima und Gesellschaft in Europa*, pp. 22–23.
[3] Rohr/Camenisch/Pribyl, "European Middle Ages"; Summerhayes, Palaeoclimatology, p. 440.

A further misinterpretation of the image was initiated when the Intergovernmental Panel on Climate Change (IPCC) in its "First Assessment Report" in 1990 used Lamb´s graph for central England with the misleading caption "Schematic diagrams of *global* temperature variations since the last thousand years".[4] This mistake was repaired in all following IPCC-assessment reports since 1996, which presented much more refined and global temperature reconstructions. They illustrated that the scale and rate of modern-day global warming significantly excels the medieval one. Furthermore, the IPCC-reports demonstrated that modern-day climate dynamics in contrast to those in medieval times cannot be explained by natural forcing only, but mainly emerge from the impact of human activities.[5]

Sceptics and deniers of anthropogenic climate change, however, remain unconvinced and prefer to use the now outdated Lamb/IPCC-graph of 1990. The same groups tend to maintain the traditional interpretation that the "Medieval Warm Period" fostered a "rise of the European civilisation" with beneficial climatic conditions, when the Vikings colonised Greenland and wine was produced even in the north of England.[6] This serves their agenda in two ways: first, to demonstrate that the scale of modern-day global warming is not exceptional in comparison with the medieval one – and can be explained by natural fluctuations in the same way. And second, to argue that warm periods in general are beneficial for human societies, as they assume the medieval one generally was. To illustrate that earlier warm periods equalled or even excelled modern-day global warming, they use versions of the graph created by Lamb. Recently, Birgit Schneider has analysed in detail the various ways by which this and other images are used and abused in the debate on global warming. She demonstrates how deniers of anthropogenic climate change manipulated such graphs to obscure the actual rate and scale of the rise of global temperatures in the last decades by cutting off the curve at an earlier point in time, for instance.[7]

Based on an increasing number of proxy data[8] recent scholarship, however, has demonstrated that the "Medieval Warm Period" between the 10th and 13th centuries CE was neither continuously warm nor "optimal" in Western and Central Europe, not to mention other parts of the globe.[9] Therefore, they mostly resorted to the term "Medieval Climate Anomaly" (MCA). It marks a period of globally higher average temperatures than the preceding "Late Antique Cold Period" and the succeeding "Little Ice Age", but with strong differences in the regional manifestation of this global climate trend, and equally interrupted by decades of lower average temperatures, which can be connected to reduced solar activity, sometimes in combination with the atmospheric impacts of major volcanic eruptions (see below).[10]

---

[4] Summerhayes, *Palaeoclimatology*, p. 442.
[5] Mathez/Smerdon, *Climate Change*, pp. 265–303; Summerhayes, *Palaeoclimatology*, pp. 442–43; Zalasiewicz/Williams, "Climate change through Earth history", pp. 59–61; Campbell, *The Great Transition*, pp. 36–38; Intergovernmental Panel on Climate Change, *Climate Change 2021*.
[6] Summerhayes, *Palaeoclimatology*, p. 440.
[7] Schneider, *Klimabilder*, pp. 235–82.
[8] For an overview on these types of data see Mathez/Smerdon, *Climate Change*, pp. 229–38, Brönnimann/Pfister/White, "Archives of Nature and Archives of Societies"; Pfister/Wanner, *Klima und Gesellschaft in Europa*, pp. 16–20 and 118–31, as well as the relevant chapters in the present volume.

## 2 Climate oscillations, proxy data and the Medieval Climate Anomaly

One factor in the temporal dynamics of the Medieval Climate Anomaly (and of global climate in general) were fluctuations in the sun´s activity, which influenced the amount of solar irradiation (as essential form of energy input) that reached planet Earth. For solar activity, which correlates with the number of observable sunspots (with a higher number of these phenomena on the sun´s surface indicating a higher amount of activity), a cyclic fluctuation of 11 years ("Schwabe Cycle") has been observed. Further cycles at a longer term (such as the "Gleissberg Cycle", between 70 and 100 years, for instance) contribute to the emergence of maxima and minima of solar activity over several decades. Systematic observations of sunspots are available from the 17th century CE onwards; for earlier periods reconstructions of solar activity are based on the analysis of concentrations of isotopes of carbon and beryllium in ice cores and other proxies. The Medieval Climate Anomaly included two solar maxima between ca. 920 and 1020 as well as between ca. 1100 and 1200/1250. Reduced solar activity (and therefore cooler global average temperatures), on the contrast, characterised the "Oort Minimum" (ca. 1010 to 1080), while the "Wolf Minimum" (ca. 1280 to 1345) already marked the transition from the Medieval Climate Anomaly to the "Little Ice Age".[11]

Large volcanic eruptions had short-term climatic effects. Eruptive ejections of aerosols caused atmospheric phenomena, which disquieted contemporary observers (such as the "dust veil" of the year 536 CE described in Greek and Latin sources[12]). They could contribute to cooler temperatures over several months due to the reduction of solar irradiation, but also other and regionally diverse climatic effects (decreased as well as increased temperatures, decreases as well as increases of precipitation). Thus, various forms of weather extremes could emerge from the atmospheric perturbations caused by volcanic eruptions. They could initiate short term climatic fluctuations also during periods otherwise characterised by higher and more stable temperature conditions as during the above-mentioned solar maxima, such as an eruption described in written sources and also identified due to its chemical signature in ice cores from Greenland for the year 939 (possibly coming from the Eldgjá on Iceland) as well as a "cluster" of eruptions (maybe in Iceland and Japan) between 1108 and 1110 (see the reconstruction of European summer temperatures in these years in **fig. 1**). Furthermore, large volcanic eruption could aggravate the effects of minima of solar activity; the huge eruption of 1257 (now attributed to Samalas on Lombok island in modern-day Indonesia) together with following volcanic events contributed on top of the incipient Wolf Solar Minimum to the transition towards the "Little Ice Age" in the late 13$^{th}$ century (see **fig. 1**).[13]

---

[9] Adamson/Nash, "Climate History of Asia (Excluding China)". For a reconstruction of European summer temperatures see for instance Luterbacher et al., "European summer temperatures".
[10] Diaz et al. "Spatial and Temporal Characteristics"; Rohr/Camenisch/Pribyl, "European Middle Ages"; Summerhayes, *Palaeoclimatology*, pp. 442–48.
[11] Usoskin, "A History of Solar Activity"; Lean, "Estimating Solar Irradiance since 850 CE", 135 and 142; Mathez/Smerdon, *Climate Change*, pp. 180–82; Summerhayes, *Palaeoclimatology*, pp. 455–64; Guiot et al., "Growing Season Temperatures"; Polovodova Asteman/Filipsson/Nordberg, "Tracing winter temperatures"; Cohen/Stanhill, "Changes in the Sun´s radiation", pp. 691–705; Dorman, "Space weather and cosmic ray effects", pp. 713–31 (with some interesting, but problematic attempts to directly correlate wheat prices in late medieval England and Europe with reconstructions of solar activity); Campbell, *The Great Transition*, pp. 37–38, 50–54.
[12] See the chapter by Mischa Meier in the present volume.

The actual regional effects of solar and volcanic climate forcing depended on their impact on regular climate oscillations between oceans and continents. For weather conditions in western Afro-Eurasia, the Northern Atlantic Oscillations (NAO) plays a decisive role. Its dynamics are measured in an index of the differences in air pressure between the Iceland low and the high over the Azores (see the map in **fig. 2**). A high difference between these air pressure regions (resulting in a positive NAO-index) usually causes warmer and wetter weather in Western and Central Europe, but drier conditions in the Mediterranean. A low difference on the contrast results in colder and drier weather in Western and Central Europe, but more humid conditions in the Mediterranean. For the more stable periods of the Medieval Climate Anomaly, such as around 950 CE or 1140 CE, a predominantly positive NAO-index was reconstructed, while weaker NAO-effects have been identified during the Oort Solar Minimum in the mid-11$^{th}$ century.[14]

In the Mediterranean, however, a further "seesaw" of precipitation conditions was observed between the west and the east of the basin, with the former often affected by more arid conditions during positive NAO-periods of the Medieval Climate Anomaly, while the latter sometimes despite the impact of the NAO experienced more humid ones. These differences can be connected to the impacts of further climate patterns on the Eastern Mediterranean: the so-called "North Sea-Caspian Pattern" (NCP) is defined by the pressure difference between the North Sea and the Caspian Sea (see **fig. 2**); a positive NCP brings cool and dry regional winter conditions to the central and eastern Mediterranean, a negative NCP warm and wet ones. Furthermore, a strong Siberian High (see **fig. 2**) could cause unusually cold airflow from Inner Eurasia to the eastern Mediterranean and the Middle East, contributing to otherwise rare phenomena such as snowfall and frost in Baghdad, which were described several times during the Oort Minimum of the 11th century (see below).[15]

---

[13] Sigl, "Timing and Climate Forcing"; Guillet, "Climatic and societal impacts"; Büntgen et al., "Cooling and societal change"; Mathez/Smerdon, *Climate Change*, pp. 176–80; Summerhayes, *Palaeoclimatology*, pp. 466–68; Stenchikov, "The role of volcanic activity"; Riede, "Doing palaeo–social volcanology"; Campbell, *The Great Transition*, pp. 55–58; Wozniak, *Naturereignisse im frühen Mittelalter*, pp. 315–19; Pfister/Wanner, *Klima und Gesellschaft in Europa*, pp. 180–82.

[14] Goosse et al., "The medieval climate anomaly in Europe"; Guiot et al., "Growing Season Temperatures"; Polovodova Asteman/Filipsson/Nordberg, "Tracing winter temperatures"; Lüning et al., "Hydroclimate in Africa"; Mathez/Smerdon, *Climate Change*, pp. 91–97; Summerhayes, *Palaeoclimatology*, pp. 437–38, 464–65; Campbell, The Great Transition, pp. 45–48; Pfister/Wanner, *Klima und Gesellschaft in Europa*, pp. 38–40.

A further oscillation pattern with even wider ranging impacts than the NAO is the El Niño-Southern Oscillation (ENSO), described as interplay between an area usually characterised by low air pressure and warm water temperatures in the western Pacific (around modern-day Indonesia) and an area of high air pressure and cooler temperatures off the western coast of South America. These "usual" conditions characterise the "neutral" state of the Southern Oscillation. The "El Niño" state (usually observed around Christmas off the coast of Peru, hence the name) is characterised by cooler than normal conditions in the western Pacific and warmer ones in the eastern Pacific. Its counterpart, "La Niña", is characterised by warmer than normal water temperatures in the western Pacific warmer and cooler ones in the eastern Pacific. These different states of the Southern Oscillation bring about significant changes in the strength of winds and the distribution of precipitation from the ocean towards the continents. As the climatologist Mathez and Smerdon explain: "El Niño commonly creates severe droughts in Australia, Indonesia, southern Africa, and the African Sahel (the southern borderland for the Sahara Desert stretching from Senegal in the west to Sudan in the east); weakens the South Asian monsoon; and brings increased winter precipitation to parts of North America. (…) La Niña similarly has significant global impacts. It causes droughts across the western United States and Mexico, brings heavy rain to northwestern Australia and Indonesia during the boreal (Northern Hemisphere) winter, and enhances the South Asian monsoon during the boreal summer."[16] For the southwest of today´s USA, recurring "megadroughts" during the 10th to 13th centuries, which have been reconstructed on the basis of tree rings and affected the Puebloan cultures (with the famous site of Mesa Verde in modern-day Colorado), have been connected with strong La Niña-events, such as around 936, 1034, 1150 and 1253 CE. A severe El Niño on the contrast in the years around 1060 CE most probably affected precipitation conditions in East Africa (via the Monsoon system, see **fig. 2**) and thereby contributed to a series of very low Nile floods in Egypt; it is equally reflected in proxy data from the Peruvian Andes.[17] Another sequence of low Nile floods around 1200 has been equally connected with a 30 years period of severe El Niños between 1180 and 1210 CE, which accompanied a general change towards more arid conditions in the Eastern Mediterranean in the late 12th and early 13th century.[18]

---

[15] Roberts et al., "Palaeolimnological evidence"; Katrantsiotis, "Eastern Mediterranean hydroclimate reconstruction"; Kushnir/Stein, "Medieval Climate in the Eastern Mediterranean".
[16] Mathez/Smerdon, *Climate Change*, pp. 71–73.

**4 The time of the Medieval Climate Anomaly in the Byzantine Empire: "Economic Expansion"…**

Thus, even across medieval Europe and the Mediterranean, not to mention other regions of the globe, nothing like a climatically coherent "Medieval Optimum" from the 10th to the 13th centuries can be identified from the palaeoclimatological data. Nevertheless, for Western and Central Europe, shorter periods (during the solar maxima, for instance) of on average more "beneficial" and stable temperature and precipitation conditions existed and were accompanied by significant demographic and economic growth.[19]

Similar phenomena of an "Economic Expansion in the Byzantine Empire" between 900 and 1200 especially since the publication of the monograph of this title by J. A. Harvey in 1989[20] (and in contrast to earlier scholarship, which linked the severe political crisis of Byzantium in the later 11th century with an assumed economic one[21]) have been identified on the basis of written sources (such as the village tax charters from Mt. Athos, for instance) and an increasing number of archaeological evidence (coin finds, settlement numbers stemming from surveys such as in Laconia or Boeotia).[22] In the fundamental "Economic History of Byzantium" in 2002, the editor Angeliki E. Laiou (1941-2008) summed up the new consensus: "Over the past few decades, the eleventh and twelfth centuries have been recognized as periods of economic growth, a judgment that goes counter to most of the earlier historiography. The main reason for the earlier perception, held by eminent historians, was that they saw Byzantium from the viewpoint of the state and considered that the military defeats and evident decline of the state in the late eleventh century, as well as the territorial retraction in the twelfth century, were paralleled by a decline in the economy. Instead, it has been recognized that, for the first time in Byzantine history, there was a disjunction between military and territorial developments on the one hand and economic activity on the other. (…) Finally, historians now look with a different eye at developments that in the past had been considered negatively: all devaluations [of the nomisma] had been thought bad, whereas now we differentiate between "devaluations of expansion" and those that result from a crisis; the large estate, once thought to signal and promote the collapse of Byzantium and its agrarian base, is now seen as a factor in economic expansion."[23]

---

[17] Mathez/Smerdon, *Climate Change*, pp. 71–88 and 252–53 on the megadroughts in North America. For the dynamics of ENSO during the medieval centuries see Campbell, *The Great Transition*, pp. 38–45, and Grove/Adamson, *El Niño in World History*, pp. 5, 20, 51–53, also on the correlation with the Nile floods. On the latter see equally Hassan, "Extreme Nile floods"; Zaroug/Eltahir/Giorgi, "Droughts and floods"; Nash et al., "African hydroclimatic variability"; Lüning et al., "Hydroclimate in Africa"; Kushnir/Stein, "Medieval Climate in the Eastern Mediterranean", 9–11; Henke/Lambert/Charman, "Was the Little Ice Age more or less El Niño–like than the Medieval Climate Anomaly".
[18] Grove/Adamson, *El Niño in World History*, pp. 52–55; Nash et al., "African hydroclimatic variability"; Hassan, "Extreme Nile floods"; Nicholson, "A Multi–Century History of Drought".
[19] Campbell, *The Great Transition*, pp. 58–86; Pfister/Wanner, *Klima und Gesellschaft in Europa*, pp. 274–84.
[20] Harvey, *Economic Expansion in the Byzantine Empire*. See also Harvey, "The Byzantine Economy in an International Context"; Preiser-Kapeller, "Byzantium 1025–1204", p. 65.
[21] See, for instance, Svoronos, *Etudes sur l'organisation intérieure*, pp. 348–50 (originally published in 1966).
[22] For an overview on the evidence especially for Greece see also Olson, *Environment and Society in Byzantium*, pp. 145–58.

Without doubt, the stabilisation of the Byzantine frontiers vis-à-vis the Arabs in Eastern Asia Minor and the Bulgarians on the Balkans (respectively the conquest of the Arab Emirate on Crete in 961) under the so-called "Macedonian" dynasty founded by Emperor Basil I (867-886) allowed for a recovery of agricultural activities.[24] But especially the last aspect mentioned by Laiou, that is the role of the "large estates" in the agricultural dynamics of this time, as well as the actual scale of the economic growth, remain controversial; in 2008. for instance, Mark Whittow (1957-2017) argued: "Rather than thinking of peasants needing to be forced onto the market and so launching an economic revolution, we should perhaps be thinking of a world where landowning aristocrats hijacked the fruits of pioneering peasant enterprise. (…) But if the economic revival of the middle Byzantine period was in fact driven from below, these vast estates might look more like a stifling of Byzantine economic enterprise. Did the great estates necessarily promote local and regional economic enterprise, or did they dampen such activity in favour of self-sufficiency and the provision of goods in kind to feed their dependents in the capital? Is the growth of Latin commerce a sign of the economic vitality of Byzantium, or rather a symptom of an economy where Byzantine traders were disadvantaged in favour of outsiders servicing the great estates?"[25]

Recently, findings from palynology (pollen data) confirmed an increase of agricultural cultivation for various regions of the Byzantine Empire from the late 9$^{th}$ century (Asia Minor) respectively 10$^{th}$ century (Southern Balkans) onwards after a significant decrease between the 6th and 8th centuries (see **fig. 4** and **fig. 5**, for the case of cerealia). They also indicate regional and local variations with regard to the increase of the cultivation of grain, olives and wine or a focus on pasturage and animal husbandry.[26] What this data cannot tell us, however, is if the expansion of agriculture in the area around the pollen sites stemmed from the initiatives of communities of small-scale peasants, large noble or monastic estates or imperial domains (or tell us about the respective shares of these forms of economic organisation in the growth).

---

[23] Laiou, "The Byzantine Economy", p. 1150. See also Kaplan/Morrisson, "L'économie byzantine: perspectives historiographiques"; Prigent, "The Mobilisation of Fiscal Resources".
[24] Kaldellis, *Streams of Gold*, p. 21; Olson, *Environment and Society in Byzantium*, pp. 144–45.
[25] Whittow, "The Middle Byzantine Economy", pp. 487–91. For an illuminating study on peasant strategies in pre–mechanized Mediterranean farming see Halstead, *Two Oxen Ahead*.

Most recently, Alexander Olson in his study on oak and olive cultivation in the middle Byzantine period has renewed the argument that "greater pressure for surplus from elite figures led to economic and demographic growth, not the other way around. Put another way, the economic upsurge and agricultural intensification were the consequences of a more demanding elite rather than demographic growth."[27] Furthermore, he identifies signs of slow agricultural growth and modest modifications of the landscape in the 10th and 11th century Aegean provinces; in his opinion, more significant economic dynamics only started in the late 11th century with "a Komnenian-led Byzantine state demanding more tax revenue, an emerging class of big landowners seeking rent, the presence of Italian merchants permitted to trade across the Aegean's waters, and peasants pressured to produce a surplus that garnered coins on the market."[28]

Surprisingly in his otherwise excellent monograph, Olson largely ignored the earlier studies of the eminent historian of medieval Mediterranean trade and economy David Jacoby (1928-2018), who in one of his last papers on the Peloponnese summed up his earlier deliberations on 11th century agricultural dynamics, which suggest a different interpretation, at least for this region: "It is commonly assumed that Italian merchants were the initiators of large-scale commercial exports of oil from the Peloponnese in the twelfth century. This proposition is rather implausible. It rests on skewed evidence – namely, the chance survival of a few Venetian charters and the absence of similar Byzantine documentation. (…) In short, there is good reason to believe that the export of foodstuffs from the Peloponnese and other provinces of the empire to Constantinople was likely initiated by large Byzantine landowners, merchants, and carriers, attentive to market demand. (…) We may safely assume, therefore, that [local archontes] encouraged the extension [of olive cultivation], both on their own demesne and on the peasants' land and that the peasants shared their market-oriented approach. The contention that Byzantine peasants produced surpluses only to pay their taxes in cash may be safely dismissed for the twelfth-century Peloponnese, even if correct for other Byzantine regions and other periods."[29] Also, the above-mentioned pollen data suggests beginnings of significant agricultural growth already in the 10th and 11th centuries (see **fig. 4** and **fig. 5**), a process intensified, but not depending on Komnenian state activity and Italian merchants to the extent suggested by Olson. Following Jacoby, we may assume a more complex interplay between the initiatives of large-, medium- and small- scale landowners and of the state (see also the illustrative example of the conflict between the monastery of Kolobou and the village of Siderokausia already of the year 995 below) – with shifting weights between these actors, however, as further discussed below. Yet, the share of climatic factors for these dynamics remains to be determined; Olson, for instance, sceptically, but – on the current basis of knowledge – legitimately states that "while I agree that climate change can have massive (even catastrophic) implications for people and plant life in various times and places, I privilege people's choices and botanical agency over climate as explanations for changes in the vegetative cover of the Aegean Basin during". [30] Nevertheless, recent studies like for Western Europe identified "beneficial" long term climatic trends for the Byzantine territories during some periods of the Medieval Climate Anomaly based on proxy data. In 2016, Elena Xoplaki and her co-authors summed up the evidence on "The Medieval Climate Anomaly and Byzantium" in a paper. They combined the above-mentioned evidence for economic and demographic growth in Byzantium between 850 and 1300 CE with new findings from palynology. Furthermore, they provided an overview on the "palaeoclimate evidence for the medieval Byzantine region", including "documentary textual evidence" (building on the pioneering work of Ioannis Telelis[31]) and "natural proxies" for the Balkans and Asia Minor. On this basis, they draw "potential links between climatic and societal changes that took place during specific periods in Byzantium", such as correlations between "a long-term trend towards wetter conditions in ca. AD 850-1000 in western Anatolia" respectively "stable and relatively warm-wet conditions in northern Greece AD 1000-1100" and periods of agricultural expansion in these regions (see the map in **fig. 3** for some proxy data sites).[32]

---

[26] Izdebski/Koloch/Słoczyński, "Exploring Byzantine and Ottoman economic history"; Olson, *Environment and Society in Byzantium*, pp. 45–46, 65–71.
[27] Olson, *Environment and Society in Byzantium*, p. 197.
[28] Olson, *Environment and Society in Byzantium*, p. 178.

---

[29] Jacoby, "Rural Exploitation", pp. 237–38.
[30] Olson, *Environment and Society in Byzantium*, pp. 13–14.
[31] Esp. the systematic catalogue of meteorological phenomena described in Byzantine and other sources in Telelis, *Μετεωρολογικά φαινόμενα*. On this basis, a list (in English) with short entries has been created for Haldon al., "The Climate and Environment of Byzantine Anatolia", 154–60. See also Telelis, "Climatic fluctuations in the Eastern Mediterranean", 167–207.
[32] Xoplaki et al., "The Medieval Climate Anomaly and Byzantium", 229–52. Also, the oxygen isotope data from Lake Nar in Cappadocia suggest a change towards a more humid trend in the mid-9th century, see Woodbridge/Roberts, "Late Holocene climate"; Dean et al., "Palaeo-seasonality of the last two millennia".

In the same volume, Alexandra Gogou and her co-authors provided additional evidence from the northern Aegean. Based on the analysis of a marine record from the Athos basin in Northern Greece, the authors reconstructed changes of sea surface temperatures (SST) and other palaeo-environmental factors. Their data suggests a cooling trend from ca. 500 to 850 CE, followed by a warming trend from ca. 850 to 950 CE in the Northern Aegean. Another positive trend of SST was identified from the end of the 11$^{th}$ century onwards, while the time from ca. 1250 to 1400 CE according to Gogou and her co-authors was characterized by "arid-like conditions", with another transition "towards more humid conditions" after ca. 1400 CE.[33]

More recent data from the southwestern Peloponnese published by Christos Katrantsiotis and his co-authors confirms these findings, with wetter conditions accompanying the transition to the Medieval Climate Anomaly in Southern Greece, with the period from 850 to 1050 reconstructed as relatively wet. The decades between ca. 1050 and 1150, however, were reconstructed as more arid.[34]

**4 … or a "Collapse of the Eastern Mediterranean"?**

This later period partly overlaps with already mentioned Oort Solar Minimum of the 11$^{th}$ century, which in general was characterised by cooler average temperatures and a higher number of weather extremes.[35] Proxy data such as the one published by Katrantsiotis as well as carbon isotope data from speleothems in the Kocain Cave in Southwestern Turkey and the Uzuntarla Cave in the European part of Turkey (see the map in **fig. 3**) indicate a general change towards cooler and more unstable conditions in the early 11$^{th}$ century.[36] In addition, written sources provide us with more or less detailed description of weather anomalies and their sometimes catastrophic impacts in the form of harvest failures, rising prices and famines, accompanied by plagues of locusts or rodents and diseases among humans and animals (see below).

Based on some of these sources, Ronnie Ellenblum (1951-2021) in 2012 published a monograph on "The Collapse of the Eastern Mediterranean", which identified climate change as prime mover for a political and economic "Decline of the East" (especially Abbasid/Buyid Mesopotamia, Fatimid Egypt and Byzantium) in the 11$^{th}$ century.[37] This counter-scenario to the above-mentioned notion of "Economic Expansion" proved quite influential until today[38], despite strong criticism by several scholars of the rather mono-causal and often insufficiently documented narrative of Ellenblum, including by myself.[39] Recently, the "collapse of the Eastern Mediterranean" has even been connected with the end of the "classical" period of Islam or an expanded "Long Late Antiquity" covering the entire first Millennium CE.[40]

---

[33] Gogou et al., "Climate variability and socio–environmental changes". See also Olson, *Environment and Society in Byzantium*, pp. 13–14.
[34] Katrantsiotis, "Eastern Mediterranean hydroclimate reconstruction".
[35] Kushnir/Stein, "Medieval Climate in the Eastern Mediterranean", 3, 11–12.
[36] Götürk, *Climate in the Eastern Mediterranean*.
[37] Ellenblum, *The Collapse of the Eastern Mediterranean.*

Ellenblum – deliberately – largely ignored the growing amount of natural scientific proxy data, which in his opinion could only record long term trends, but not severe short-term shocks and supply short falls which were only described in the written sources (what Christian Pfister has called the "archives of society"[41]) and in his opinion contributed to the destabilisation of regimes in Baghdad, Cairo or Constantinople or the collapse of infrastructures and communities.[42] Ellenblum´s criticism is legitimate to a certain extent; several types of proxies such as pollen data or lake sediments often allow only for a reconstruction of dynamics at a time scale of centuries or several decades.[43] And even if tree rings or fine graded sediments enable annual reconstructions of temperature or precipitation conditions, their spatial coverage (often limited to the environs of the site from where the data is retrieved) or the seasonality of their signal (tree rings, depending on latitude and altitude, may indicate precipitation mainly for the spring, for instance) do not allow for a correlation with weather extremes described in written sources for other regions or times of the year.[44] However, also the corpus of historiographical evidence is characterised by an highly unequal temporal and especially spatial resolution, so that due to the observations points of authors we have much more information on extreme events for the capitals of Baghdad, Cairo and Constantinople and their environs than for other regions of the Abbasid/Buyid, Fatimid or Byzantine empires.[45] Furthermore, descriptions of meteorological phenomena (in the tradition of antiquity including not only weather conditions but equally other celestial signs such as comets, meteors or eclipses[46]) as well as epidemics and famines were included into histories and chronicles never as mere factual reports, but always serve narrative purposes. Often, they stem from a "moral meteorology", which attributes portents and catastrophes to the flaws and misdoings of individuals rulers, elites, or societies at large, of the past or present.[47] In addition, they were interpreted as harbingers of a king´s fall or death or even the imminent end of the world (in the apocalyptic traditions of Judaism, Christianity, or Islam). Especially the late 10th and early 11th century with the turn of the first Millennium since the birth of Jesus Christ, the 1000th anniversary of his crucifixion and resurrection (dated to around 1030) and other significant dates identified by studious calculators provided several anchor points for such interpretative frameworks. Explicit in this regard are the historical works of the Byzantine Leo the Deacon (late 10th century) or the Armenian histories of Aristakēs Lastiverc'i (11th century) and Matthew of Edessa (12th century, see below).[48] The end time-atmosphere of their texts may have contributed to Ellenblum´s development of the grim scenario of a "Collapse of the Eastern Mediterranean" in the 11th century.[49]

---

[38] See for instance Kushnir/Stein, "Medieval Climate in the Eastern Mediterranean", 11–2 (the authors refer to my earlier study ["A Collapse of the Eastern Mediterranean?"], which critically re–evaluated Ellenblum´s book, on 20, n. 30, but did not take into consideration its findings), or Wozniak, *Naturereignisse im frühen Mittelalter*, pp. 493–94.

[39] The review on Ellenblum´s book by S. White in: *Mediterranean Historical Review* 28 (2013), 70–72. Cf. also the review by S. Harris in: *Journal of Historical Geography* 42 (2013), 218–19, and especially Preiser-Kapeller, "A Collapse of the Eastern Mediterranean?". See also Kaldellis, *Streams of Gold*, p. xxxi, who states: "Recent efforts to write narrative history from a climatological angle seem to be reductive and fail to explain Byzantine expansion in an age of supposed regional collapse".

[40] Fowden, *Before and after Muḥammad*; Bauer, *Warum es kein islamisches Mittelalter gab*.

[41] On the value of written evidence for the reconstruction of climates of the past and the problems of their interpretation see Brönnimann/Pfister/White, "Archives of Nature and Archives of Societies"; Pfister/Wanner, *Klima und Gesellschaft in Europa*, pp. 131–42; Nash, "Climate indices".

[42] See also the posthumous publication Chipman/Avni/Ellenblum, "Collapse, affluence, and collapse", where it is stated already in the abstract: "Well–dated historical sources are the only way to follow the climatic and societal occurrences in a yearly resolution. No proxy data are sensitive enough to detect such changes and to reconstruct the historical and social processes that followed the climatic anomalies."

[43] Izdebski/Koloch/Słoczyński, "Exploring Byzantine and Ottoman economic history".

[44] Xoplaki et al., "Modelling Climate and Societal Resilience", for the unequal spatial distribution of proxies in the Eastern Mediterranean.

Furthermore, both the Byzantine Empire and the Fatimid Caliphate survived the 11th century, albeit on a modified and reduced power base; the written sources not only provide narratives of doom, but also hint at the resilience of societies and regimes, which equally could use crisis management as opportunity to stabilise support and increase their legitimacy, at least from a symbolic point of view (see below). In addition, we have to contrast these short-term shocks with the undoubted long term economic growth in 10th to 11th centuries Byzantium and Egypt as documented in other forms of written sources (such as charters), archaeological evidence and the increasing number of proxies such as pollen.[50]

---

[45] Grotzfeld, "Klimageschichte des Vorderen Orients"; Vogt et al., "Assessing the Medieval Climate Anomaly in the Middle East"; Xoplaki et al. "The Medieval Climate Anomaly and Byzantium", 239.
[46] Dagron, "Les diseurs d'événements"; Magdalino, "Astrology"; Tihon, "Astronomy".
[47] For this term see Elvin, "Who Was Responsible for the Weather". See also Burman, "The political ontology of climate change"; Pfister/Wanner, *Klima und Gesellschaft in Europa*, pp. 96–99.
[48] Wozniak, *Naturereignisse im frühen Mittelalter*, pp. 715–31; Neville, *Guide to Byzantine Historical Writing*, pp. 124–25; Brandes, "Byzantine Predictions of the End of the World"; Andrews, *Mattʿēos Uṙhayecʿi and His Chronicle*.
[49] Ellenblum, *The Collapse of the Eastern Mediterranean*, p. 129.
[50] On evidence of economic growth in Fatimid Egypt see Frantz-Murphy, "A New Interpretation of the Economic History of Medieval Egypt"; Frantz-Murphy, "Land Tenure in Egypt"; Mayerson, "The Role of Flax in Roman and Fatimid Egypt"; Udovitch, "International Trade"; Chipman/Avni/Ellenblum, "Collapse, affluence, and collapse", 205–09. For Byzantium, see the evidence described above.

In the following sections, we follow Ellenblum´s lead to focus on the written sources for the 10th and 11th centuries, used (or more often not used) by Ellenblum, whose short chapter on Byzantium is relatively poorly documented when it comes to Byzantine historiography, with a few passages from John Scylitzes, Scylitzes Continuatus, Michael Psellos, Michael Attaleiates and Nicephorus Bryennios. Furthermore, he concentrated on the wars with the Pechenegs and Oghuz, so that not a single description of a climatic phenomenon from a Byzantine source is found in his chapter; characteristically, the few meteorological passages cited he took only from Matthew of Edessa.[51]

In addition to a close reading of relevant passages from Byzantine historiography, we try to find additional contemporary observations on climatic extremes from neighbouring regions of the Eastern Mediterranean and Middle East (with Telelis´ ground-breaking catalogue serving as indispensable guide once more[52]), but also Western Europe and China. Furthermore, historiography is augmented with information from imperial legislation and charters as well as proxy data, if existing and significant. Thereby, we aim to put into perspective both Ellenblum´s "collapse" as well as Harveys "economic expansion" through a more nuanced reconstruction of the short- and long-term results of the interplay between environmental and social dynamics.

**5 The winter of 927 as short-term shock and the long-term dynamics of economic growth and socio-political changes in 10th to 11th century Byzantium**

As mentioned above, already during the more stable periods of the Medieval Climate Anomaly such as the solar maximum between ca. 920 and 1000, short term extremes left their traces in the sources. An often-discussed case is the "extreme winter" of 927/928 CE. Already before, after the death of Tsar Simeon on 27 May 927, "the Bulgar nation was suffering a severe famine and a plague of locusts which was ravaging and depleting both the population and the crops. (…)".[53] Then, in December 927, according to John Skylitzes (written in the late 11th century[54]), "an intolerable winter suddenly set in; the earth was frozen for one hundred and twenty days. A cruel famine followed the winter, worse than any previous famine, and so many people died from the famine that the living were insufficient to bury the dead. This happened although the emperor [Romanos I Lakapenos] did his very best to relieve the situation, assuaging the ravages of the winter and the famine with good works and other aid of every kind."[55]

---

[51] Ellenblum, *The Collapse of the Eastern Mediterranean*, pp. 123–46.
[52] Telelis, Μετεωρολογικά φαινόμενα. Ellenblum unfortunately did not use this systematic survey.
[53] John Scylitzes, *Synopsis*, 18, ed. Thurn, pp. 222, 18–20; transl. Wortley, pp. 215. Cf. also Telelis, Μετεωρολογικά φαινόμενα, nr 372. A chronological link of such calamities to the death of a ruler, however, is often used by medieval historians to underline the severity of the loss, see Wozniak, *Naturereignisse im frühen Mittelalter*, p. 576.
[54] Neville, *Guide to Byzantine Historical Writing*, pp. 155–57.

The winter of 927/928 – along the lines of Ellenblum´s argument – illustrates the limits of the "archives of nature"; as Adam Izdebski, Lee Mordechai and Sam White have pointed out, "no currently available palaeoclimate proxy confirms the occurrence of a strong winter cooling, or at least a potential increase in snowfall" exactly at the time of the winter of famine.[56] Isotopes analyses for the Nar Gölü in Cappadocia (see the map in **fig. 3**) only hint at on average more snow-rich winters for the period from 921 to 1071 CE (which would also include the extreme winter of 927/928). Similarly, a long-term cooling trend can be observed in speleothem-data from the Kocain Cave in south-western Anatolia (see **fig. 3**).[57] We have, however, another observation on the severity of the winter 927/928 in the Chronography of Bar Hebraeus for Baghdad, who wrote that "there was a very bitterly cold winter again in Baghdad. It was so cold that the vinegar of the wine [the freezing point of five percent wine vinegar is -2 degrees Celsius] in the cellars and eggs, and oils, froze, and the trees withered".[58] Latin sources speak about a "very great winter" in the Rhine region.[59]

Further information on the relieve measures of Romanos I Lakapenos during the extreme winter can be found in the Chronicle attributed to Simon Logothetes (written in the 960s[60]): the emperor ordered the installation of wooden emergency shelters in the numerous porticos along the main streets of Constantinople to save the homeless from death by freezing; furthermore, money was collected in the churches for the needs of the poor, some of whom were also regularly feed in the imperial palace.[61]

Equal praise of the posteriority Romanos I earned with another measure, which tackled a longer term impact of the winter of famine seven years later: the novel of September 934 (with a probable first version issued already in 928/929[62]), which disposed that "from the previous first indiction (that is, from the advent or passage of the famine [during the winter of 927/928]), those illustrious persons, whom the present decree prescribes above be prohibited, who have

---

come into control of hamlets or villages and have there acquired further properties either in part or in whole, are to be thence expelled, recovering the price they paid either from the original owners or from their heirs or relatives, or if these persons do not have the means, from the joint taxpayers, or even from the commune coming forward to return the price."[63] This law was directed against those among the "powerful" (*dynatoi*)[64], who as owners or administrators of large estates "regard the poor as prey" and had abused their "indigence" as "opportunity for business instead of charity, compassion, or kindness"; and "and when they saw the poor oppressed by famine, they bought up the possessions of the unfortunate poor at a very low price, some with silver, some with gold, and others with grain or other forms of payment."[65] Such distress sales under the impact of the extreme winter of 927/928 and during the following years were to be annulled in order to restore small- and medium sized farmers to their property. For their wellbeing, the law decreed, "demonstrates the great benefit of its function — the contribution of taxes and the fulfilment of military obligations — which will be completely lost should the common people disappear. Those concerned with the stability of the state must eliminate the cause of disturbance, expel what is harmful, and support the common good."[66] Already earlier laws had dealt with the growing influence of the *dynatoi* at the cost of the small free farmers, some of whom not only provided taxes, but also military service (the owner of *stratiotika ktemata*) to the state.[67] As Angeliki Laiou and Cecile Morrisson have pointed out, this process was also carried by dynamics from below, that is farmers preferring to sell their land to powerful elite neighbours and to work for them as tenants in order to profit from their protection and resources in cases of political unrest or – as in the aftermath of 927 – of crop failures and extreme events.[68] We may assume that this long-term trend started with the improvement of security and economic (as well as climatic) conditions in the 9th century, intensified with the imperial and economic expansion of the 10th and early 11th century, and heated up in aftermath of short-term shocks such as in 927. Besides its actual socio-economic impact, the extreme winter of 927/928 (which also served as reference point in later laws repeating and expending the regulations of Romanos I[69]) "was thus used as a way of understanding the reasons for the social transformation as it accelerated this process and brought it to contemporaries' attention."[70] In general, as Adam Izdebski, Lee Mordechai and Sam White have analysed, "the environmental stressor [of 927/928] – even if in physical terms it was not the harshest winter of the tenth century – impinged upon the complex web of crop ecologies, social relations, and the state's interests. Thus, it added new momentum to the extant social dynamic – that of officeholding elites accumulating wealth that allowed them to become an increasingly powerful social group within contemporary Byzantine society."[71] Furthermore, "from a broad perspective, (…) Byzantine society proved resilient, surviving the crisis caused by the long winter of 927/928. When seen from the point of view of specific social groups, however, the price for this resilience was a significant shift in the balance of socioeconomic relations."[72]

---

[63] Novels of the Macedonian emperors, ed. Svoronos, nr. 3, p. 86, lns. 88–95; transl. McGeer, *The Land Legislation of the Macedonian Emperors*, p. 56; Dölger/Müller/Beihammer, *Regesten von 867–1025*, nr 628. See also Kresten, ’Ἄρκλαι und τριμίσια", 40–41; Kaplan, *Les hommes et la terre*, pp. 421–25.
[64] The law lists "illustrious magistroi or patrikioi, […] any of the persons honoured with offices, governorships, […] civil or military dignities, […] anyone […] enumerated in the Senate, […] officials or ex–officials of the themes, […] metropolitans most devoted to God, archbishops, bishops, abbots, ecclesiastical officials, or supervisors and heads of pious or imperial houses", Novels of the Macedonian emperors, ed. Svoronos, nr. 3, p. 84, lns. 50–56; transl. McGeer, *The Land Legislation of the Macedonian Emperors*, pp. 54–55; Morris, "The Powerful and the Poor", 14.
[65] Novels of the Macedonian emperors, ed. Svoronos, nr. 3, pp. 83–86; transl. McGeer, *The Land Legislation of the Macedonian Emperors*, pp. 54–56.
[66] Novels of the Macedonian emperors, ed. Svoronos, nr 3, p. 85, lns. 69–74; transl. McGeer, *The Land Legislation of the Macedonian Emperors*, p. 55.
[67] Dölger/Müller/Beihammer, *Regesten von 867–1025*, nr 595; Kaldellis, *Streams of Gold*, pp. 10–11.
[68] Laiou/Morrisson, *The Byzantine Economy*, p. 106. See also Kaplan, *Les hommes et la terre*, pp. 391–97.

Most recently, however, Anthony Kaldellis has put into question such a "significant shift" in the balance of socioeconomic relations and the traditional scenario of a conflict between landowning "magnates" and the imperial centre. He on the contrast claims, "the emperors were threatened not by landowners but by army officers. Some were no doubt landowners, but there is no evidence that they were dangerous because of their property. (…) Instead, they were dangerous because they could subvert the loyalty of the armies."[73] Kaldellis thus interprets the crisis of the 11th century as "systemic crisis" of the usual (and usually often fragile) power arrangement of emperor, army and state apparatus.

As a matter of fact, beyond the apparent support for the independence of farmers and especially soldier-farmers, a main aim of the laws of Romanos I Lakapenos and his successors was the (re-)intensification of the access of the fiscus onto the fruits of the economic growth in the countryside. The same purpose served confiscations (also under the pretext of these new laws), reclaims of abandoned lands earlier made accessible for free to neighbouring peasants, and the creation of imperial domains in old and in newly-conquered territories.[74] Furthermore, a close reading of jurisdiction of the 10th and early 11th century (such as the *Peira*) indicates that this new legislation was enforced also after the last of these "Macedonian" laws was issued by Emperor Basil II in 996.[75] Thus, Kaldellis is right to point out the still strong institutional muscles of the Roman state against earlier notions of a "feudalisation" of Byzantium. He claims that all significant means of power were still located within the imperial system and state apparatus and that the so-called "magnates of Asia Minor" were not able to challenge the emperor in Constantinople based on their new landed properties, but only due to their functions and prestige within the army and/or administration.[76]

---

[69] Dölger/Müller/Beihammer, *Regesten von 867–1025*, nr 656 (March 947), nr 707e (966/967) nr 783 (January 996). See also Morris, "The Powerful and the Poor", 5–6.
[70] Izdebski/Mordechai/White, "The Social Burden of Resilience". See also Morris, "The Powerful and the Poor", 8–10; Sarris, "Large Estates and the Peasantry", 446–47.
[71] Izdebski/Mordechai/White, "The Social Burden of Resilience".
[72] Izdebski/Mordechai/White, "The Social Burden of Resilience".
[73] Kaldellis, *Streams of Gold*, pp. 13–15, also 224–28. For a similar interpretation also for elite rebellions in the late 12th century see Olson, *Environment and Society in Byzantium*, pp. 209–10.
[74] Sarris, "Large Estates and the Peasantry", 433–34; Kaldellis, *Streams of Gold*, pp. 147–48.

A further reading of the laws of the Macedonian emperors, however, allows us to observe the actual interplay between the traditional allocation of rank and wealth within the army and administration and the growth of landed property and their transmission within increasingly powerful noble clans. In 996 CE, Emperor Basil II issued a law which he claimed to have based on his findings of conditions in the Anatolian provinces when travelling there. The emperor described the case of man called "Philokales, who was originally one of the poor and the villagers, but afterwards one of the illustrious and wealthy, and who as long as he was among the lowly paid his taxes with his fellow villagers and did not interfere with them; but when God raised him to the title of *hebdomadarios*, then *koitonites*, and thereafter *protobestiarios*, in short order he took possession of the entire village community and made it into his own estate; he even changed the name of the village."[77] The promotion in the administration of the state thus provided Philokales with the means and influence in order to expand his landed property and to turn around completely the balance of power in his home region to his advantage. Furthermore, Emperor Basil II narrates how this interplay between rank and influence in the state apparatus and growing wealth and influence in the countryside worked even more for those "well-born" clans able to maintain and to transmit this combination of public office and private enrichment over generations ("for one hundred or even one hundred and twenty years"), such as the Phokas-family.[78]

---

[75] Morris, "The Powerful and the Poor in Tenth–Century Byzantium", 27; Magdalino, "Justice and Finance in the Byzantine State"; Gkoutzioukostas, "Administrative structures of Byzantium"; Prigent, "The Mobilisation of Fiscal Resources"; Dölger/Müller/Beihammer, *Regesten von 867–1025*, nr 688a (after October 960). This runs against Kaldellis´ assumption that the laws of the Macedonian emperors were primarily "rhetoric" or "symbolic", see Kaldellis, *Streams of Gold*, pp. 17–18, 118–19. On the legislation of the Macedonian emperors see also Schminck,"Zur Einzelgesetzgebung".

[76] Kaldellis, *Streams of Gold*, pp. 13–18.

[77] Novels of the Macedonian emperors, ed. Svoronos, nr 14, p. 203, lns. 51–58; transl. McGeer, *The Land Legislation of the Macedonian Emperors*, p. 119. Cf. Dölger/Müller/Beihammer, *Regesten von 867–1025*, nr. 783; Morris, "The Powerful and the Poor", 13–14.

[78] Novels of the Macedonian emperors, ed. Svoronos, nr. 14, p. 203, lns. 38–45; transl. McGeer, *The Land Legislation of the Macedonian Emperors*, p. 117. Cf. also Morris, "The Powerful and the Poor", 16–17; Whittow, "The Middle Byzantine Economy"; Cooper/Decker, *Life and Society in Byzantine Cappadocia*; Grünbart, *Inszenierung und Repräsentation*; Andriollo, *Constantinople et les provinces d'Asie Mineure*; Andriollo/Métivier, "Quel rôle pour les province"; Sarris, "Large Estates and the Peasantry", 436–37.

Even more efficient regarding the long-term transmission of wealth and privilege were monasteries, whose institutional arrangements did not depend on the fortunes of genealogy as lay families did from one generation to the next; they equally tended to infringe upon the customary rights of villages adjacent to their domains. In an early example of December 995 (also from the reign of Basil II), the monks of Kolobou (founded in 866 and later in 980 obtained by the Monastery of Iviron on Mt. Athos) near Hierissos tried to impede the inhabitants of the nearby village of Siderokausia from further usage of an area called Arsinikeia, where the villagers had been accustomed to pasture their animals, gather wood, harvest chestnuts and plant crops. Furthermore, the monastery´s cattle had damaged the crops in the area of Kato Arsinikeia which the villagers had cultivated ("While Lower Arsinikeia, as mentioned, was once completely covered in woodland and trees, with canals dug to bring waters from the heights to operate the mills and make fertile the gardens and orchards, as well as the grazing lands for the beasts, the monks thought it would be a good idea to allow a multitude of animals loose there, which ruined the seeds that the Siderokausites living there had planted"). The (informal) customary usage and significant modification of the landscape by a village community collided with the legal claims of an expanding monastic landowner. In this case, the judge decreed a division of the usage of the areas around Arsinikeia between Siderokausia and Kolobou, but in many similar cases, monastic institutions would be even more successful in enforcing their interests.[79]

For the lay families, however, Kaldellis is right again in his observation on the general fragility of elite status in Byzantium; in contrast to the Western Empire, noble status and access to offices never became hereditary.[80] As John Haldon points out: "Being a member of (Byzantine) elite was never (…) a fixed or determinate quantity – on the contrast, it was to occupy a position in a complex set of social and cultural relationships, in which position, status, and income remained both negotiable and fragile."[81] Those families able to transmit wealth and power across generations "for one hundred or even one hundred and twenty years" (as Basil II has claimed) remained exceptional. In a statistical analysis of the material on elite families of the 10th to the 12th centuries collected in the study of Alexander Kazhdan and Silvia Ronchey, we observe a significant "turnover" of elite families during this period; out of 282 families in the sample, only four were able to maintain their status throughout the entire 228 years covered in the study and only ca. 15 % for more than 100 years. The vast majority lost elite status again after three, two or even only one generation, with the 11[th] century showing an especially high "turnover".[82]

---

[79] *Actes d'Iviron* I, ed. Lefort/Oikonomidès/Papachryssanthou, nr 9, pp. 160–63. For a detailed discussion and English translation of this charter, see now Kaplan, "The Monasteries of Athos and Chalkidiki", 72–77. See also Olson, *Environment and Society in Byzantium*, pp. 215–216 and 219 (who correctly remarks that "appropriation of a resource can generate as much societal pressure as resource depletion"); Smyrlis/Banev/Konstantinidis, "Mount Athos", pp. 41–42.
[80] Kaldellis, *Streams of Gold*, pp. 3–4. Cf. also Barthélemy, "L´aristocracie franque".
[81] Haldon, "Social élites, wealth, and power", p. 174.

The potential effect of the fragility of elite status on political stability, however, may have been counterintuitive. As again John Haldon has pointed out, especially since the territorial shrinking of the Byzantine Empire in the 7[th] century due to the Arab expansion, Constantinople and the imperial apparatus even more than before had become the source of wealth and prestige. Thus, even more aspiring elite members and their descendants had to guarantee their access to either imperial favour or the imperial office itself, which increased the competition for the throne in Constantinople.[83] When the improved security situation and the territorial expansion of the empire from the late 9[th] century onwards allowed the "powerful" to entangle their revenues and prestige from the state with extension of their landed properties and influence in the countryside, there was even more to lose when a family lost access to the imperial system. With the inclusion of leading members of the office-holding and landowning elite at time of minority of the heir to the throne of the Macedonian dynasty, starting with Romanos I Lakapenos (who ruled for Constantine VII) and continuing with Nikephoros II Phokas and John I Tzimiskes (for Basil II and Constantine VIII), these clans had also tasted from the imperial office and were eager to maintain their grasp on it.[84]

As I have argued elsewhere, these observation correlate with more general theories from the field of institutional economic which suggest that also periods of economic growth can destabilize earlier elite arrangements due to the concomitant shift in the balances of power between factions, clans as well as centres and peripheries. The long-term trend in the transformation of the Byzantine countryside, partly supported by periods of more stable and "beneficial" climatic conditions", but also intensified by short-term shock such as the winter of 927 (and further "cluster of calamities", which we inspect on the following pages), thus contributed to an increased competition for the imperial office in the Byzantine centre, which weakened its abilities to react to exogenous challenges, and a decrease of the centre´s integrative power at the peripheries.[85]

**6 A cluster of calamities in the 960s and regime changes in Constantinople and Egypt**

---

[82] Calculations based on data from Kazhdan/Ronchey, *L'aristocrazia bizantina*.
[83] Haldon, "Social élites, wealth, and power", p. 177.
[84] Kaldellis, *Streams of Gold*, pp. 21–23, 42–44; Preiser-Kapeller, "Byzantium 1025–1204", pp. 65–66.
[85] Preiser–Kapeller, *Der Lange Sommer und die Kleine Eiszeit*, pp. 128–29, 178–81.

During the reign of Romanos I Lakapenos (920-944), another natural disaster seems to have started which plagued the empire for the following years until the reign of his namesake and grandson Romanos II (959-963) in the form of an epidemic of cattle:

> "In those days [in the reign of Romanos II] the cattle disease was raging which had plagued the Roman empire for some time, a disease known as *krabra* that wastes and destroys bovines. They say that it originated in the days of Romanos [I] the Elder. It is said that when he was constructing a palace in which gain relief from the summer heat close to the cistern of Bonos [in Constantinople], the head of a marble ox was found while the foundations were being dug. Those who found it smashed it up and threw it into the lime kiln; and from that time until this there was no interruption in the destruction of the bovine race in any land that was under Roman rule."[86]

Tim Newfield has collected parallel reports on an epizootic among cattle from Western Europe between 940 and 944 and suggests a possible connection of the outbreak and spread of the disease with climate anomalies in the aftermath of volcanic eruptions in 939/940 such as the one of the Eldgjá on Iceland.[87]

In the early reign of Romanos II for October 960 equally a lack of grain and an increase of prices is reported for Constantinople, which the emperor tried to mitigate with the purchase of grain in "west and east".[88] Weather extremes may have contributed to this shortfall; we read about unusual cold and heavy rains during the (ultimately successful) Byzantine expedition against Arab Crete in 960/961.[89] In January 960, a lack of rain afflicted Baghdad (the 10th century in general is marked as on average more arid than preceding and following periods in the oxygen isotope data from the speleothems in Kuna Ba Cave in Northern Iraq, with the years between 960 and 986 as the driest in the second half of the 10th century, see **fig. 6**); a plague of locusts devastated the countryside of Iraq in March. In February 961, a strong hailstorm beset Baghdad, as it did in the following winter of 962/963.[90] In 963/964, "there was a great famine in Cilicia [which also impeded some of the Byzantine campaigns in the area], and a great many of the people of the Arabs left and fled to Damascus. And there was also a severe famine in Aleppo, and in Harran and in Edessa".[91] In 963 and 964, also parts of Italy were affected by famine, while severe floods affected the provinces along the Yellow River in China between 964 and 968.[92]

---

[86] John Scylitzes, *Synopsis*, Romanos II, 8, ed. Thurn, pp. 251–52; transl. Wortley, pp. 242–43. Schminck,"Zur Einzelgesetzgebung", 281, note 73 (arguing of a dating of the lack of grain to October 961). For further evidence see also PmbZ nr 26834, note 17.
[87] Newfield, "Early Medieval Epizootics"; Newfield, "Domesticates, disease and climate". See also Sigl, "Timing and Climate Forcing"; Wozniak, *Naturereignisse im frühen Mittelalter*, pp. 680–81. For references to climate extremes in the 940s in Egypt and Mesopotamia see also Telelis, Μετεωρολογικά φαινόμενα, nr 376 and 377.
[88] Theophanes Continuatus VI, 13, ed. Bekker, p. 479, 1–11. Telelis, Μετεωρολογικά φαινόμενα, nr 394; Teall, "The Grain Supply"; Kaldellis, *Streams of Gold*, p. 32. For references to extreme events and famines in neighbouring regions of Byzantium such as Armenia and Mesopotamia during the 950s see Matthew of Edessa, *History* I, 1, transl. Dostourian, p. 19; al-Maqrīzī, *Ighāthah*, transl. Allouche, p. 29; Bar Hebraeus, *Chronography*, transl. Budge, p. 165 and 167; Telelis, Μετεωρολογικά φαινόμενα, nr 388 and 391–393.
[89] Telelis, Μετεωρολογικά φαινόμενα, nr 395; Kaldellis, *Streams of Gold*, pp. 34–38.
[90] Busse, *Chalif und Großkönig*, p. 387; Telelis, Μετεωρολογικά φαινόμενα, nr 396, 397. For the Kuna Ba data see Sinha et al., "Role of climate".

Equally, in 963 started a series of low Nile floods, which would continue until 969 and contributed to the downfall of the dynasty of the Iḫšīdiyūn, which had ruled over Egypt and Syria (more or less) independently from the Abbasid Caliph in Baghdad since 935. Especially, when Abū l-Misk Kāfūr, vizir and since 946 de facto ruler of the realm, died in 968, "unrest increased and riots multiplied. Much strife between the soldiery and the commanders resulted in a great loss of human life. Markets were looted and several buildings were burned. The fear of the populace intensified: they lost their wealth and their spirits. Prices became high and it was difficult to find foodstuffs, to the point that one measure of wheat sold for one dinar."[93] Eventually, in 969 an army of the Fatimid Caliph al-Muʿizz li-Dīn Allāh from North Africa under the command of Ǧauhar aṣ-Ṣiqillī invaded Egypt and brought the Iḫšīdiyūn regime to an end. With a strict regulation of the grain market, Ǧauhar mitigated the distress in al-Fusṭāṭ, which together with a return of adequate Nile floods from 971 onwards contributed to the acceptance of the new Fatimid rule in the country, which manifested itself in the building of a new residential city to the north of al-Fusṭāṭ, al-Qāhira ("the Conquering"), i. e. Cairo.[94]

The turbulences in the Iḫšīdiyūn territories during these years (together with the death of the Emir of Aleppo, Saif ad-Daula of the Hamdanid-dynasty, in 967) also eased the Byzantine expansion towards Cilicia (to which an Egyptian fleet was still sent in 965) and Northern Syria under Emperor Nikephoros II Phokas (963-969), whose troops conquered Antioch in October 969, while severe rainfalls had enforced an abandonment of an earlier siege in December 966.[95] Yet also Byzantium was perturbed by portents and catastrophes, at least according to the apocalyptically inspired history of Leo the Deacon, who mentions an earthquake in northwestern Asia Minor in 967 and a severe storm and flooding in Constantinople and its environs on 21 June of the same year, so that "people wailed and lamented piteously, fearing that a flood like that fabled one of old was again befalling them. But compassionate Providence, which loves mankind, thrust a rainbow through the clouds, and with its rays dispersed the gloomy rain, and the structure of nature returned again to its previous condition. It so happened that there was a later downpour, which was turbid and mixed with ashes (*tephra*), as in the soot from a furnace, and it seemed lukewarm to those who touched it."[96] Furthermore, on 22 December 968 "an eclipse of the sun took place", so that once more "people were terrified at the novel and unaccustomed sight, and propitiated the divinity with supplications, as was fitting". As Leo does not forget to mention, he was an eyewitness, since "at that time I myself was living in Byzantium, pursuing my general education."[97]

---

[91] Bar Hebraeus, *Chronography*, transl. Budge, p. 170. Telelis, *Μετεωρολογικά φαινόμενα*, nr 398; Hassan, "Extreme Nile floods".
[92] Wozniak, *Naturereignisse im frühen Mittelalter*, p. 615 (with citation of sources); Kaldellis, *Streams of Gold*, pp. 46–47; Zhang, *The River, the Plain, and the State*, pp. 110–12; Mostern, *The Yellow River*, pp. 123–25.
[93] al-Maqrīzī, *Ighāthah*, transl. Allouche, pp. 30–31. Telelis, *Μετεωρολογικά φαινόμενα*, nr 399. See also Chipman/Avni/Ellenblum, "Collapse, affluence, and collapse", 202.
[94] al-Maqrīzī, *Ighāthah*, transl. Allouche, pp. 30–31. Telelis, *Μετεωρολογικά φαινόμενα*, nr 399.
[95] Telelis, *Μετεωρολογικά φαινόμενα*, nr 401; Kaldellis, *Streams of Gold*, pp. 28–29, 38–40, 46–49, 60–62.

In addition to these short term portents and calamities, a shortage of grain of three or even five years duration reportedly troubled the population during the reign of Nikephoros II Phokas (963-969).[98] The causes for this famine are not mentioned, but in the History of John Scylitzes we are informed that in May 968 "there were fierce, burning winds (…), which destroyed the crops, even the vines and trees, with the result that in the twelfth year of the indiction there was an intense famine."[99] A change towards more arid conditions in the late 960s, which continued until the early 11th century, is also indicated in the isotope data from the speleothems in the Sofular cave in Northwestern Asia Minor (see **fig. 7**).[100] In 968, also Iraq was affected by drought and famine.[101]

In the following passage, John Scylitzes wrote that "the emperor (who ought to have been concerned for his subjects' well-being) now shabbily sold the imperial grain, profiting from the misfortune of those in need, and he congratulated himself as though it were some great deed that, when [grain] was selling for one gold piece a bushel, he ordered it to be sold for two". Only when Nikephoros II learned about the actual dimension of the shortage, "he immediately put the imperial and public grain [reserves] on the market, stipulating that twelve [bushels] were to be sold for one piece of gold. God approved of this good work and provided great abundance for men". This measure, however, according to Scylitzes did not temper the bad reputation the emperor had acquired due to his earlier stinginess (and also a reduction of the weight of the gold coins).[102] The low popularity of the emperor according to some modern scholars may also "partly explain the somewhat feeble public reaction to the murder of Nikephoros" by his relative John I Tzimiskes in December 969.[103] The new emperor "put an end to the relentless evil of famine by the importation of grain, which he collected quickly [and] with forethought from markets everywhere, stopping the spread of such a calamity"[104] – and thereby gaining popularity.

---

[96] Leo the Deacon, *History* IV, 9, ed. Hase, pp. 69–70; transl. Talbot/Sullivan, pp. 117–119. Telelis, *Μετεωρολογικά φαινόμενα*, nr 402; Wozniak, *Naturereignisse im frühen Mittelalter*, pp. 287–88 (on the earthquakes). The ashes could have been the result of an eruption of Vesuvius in 968, see Wozniak, *Naturereignisse im frühen Mittelalter*, pp. 329–30.
[97] Leo the Deacon, *History* IV, 11, ed. Hase, p. 72; transl. Talbot/Sullivan, pp. 122–23. Telelis, *Μετεωρολογικά φαινόμενα*, nr 402; Wozniak, *Naturereignisse im frühen Mittelalter*, pp. 196, 216–17.
[98] John Scylitzes, *Synopsis*, John I, 3, ed. Thurn, pp. 286, 48–56; transl. Wortley, pp. 273–74; Leo the Deacon, *History* VI, 8, ed. Hase, p. 103; transl. Talbot/Sullivan, pp. 152–53. Telelis, *Μετεωρολογικά φαινόμενα*, nr 406.
[99] John Scylitzes, *Synopsis*, Nikephoros II, 20, ed. Thurn, pp. 277, 37–43; transl. Wortley, p. 266.
[100] Fleitmann et al., "Sofular Cave".
[101] Busse, *Chalif und Großkönig*, p. 387.

Thus, the end of the famine crises of the 960s in 969[105] both in Egypt and in Byzantium was accompanied by a violent regime change, with the new rulers legitimising their takeovers by more efficient crisis management. In turn, the sources suggest that the severity of the supply shortfall did not only depend on the strength of the environmental stressors, but equally on the decision-making of the earlier administrations. Thus, "climate" on its own did not topple the Iḫšīdiyūn or Nikephoros II. Nevertheless, the obviously tense economic situation may have contributed to a further intensification of the growth of large estates of the "powerful" in the Byzantine provinces; at least, Nikephoros II issued another five laws repeating and augmenting the legislation against the encroachments of *dynatoi* on the properties of their poorer neighbours. However, as mentioned above, these novels equally provided pretexts for the state to intervene in the (re-)distribution of the increasing pool of economic resources in the countryside for its very own material interests.[106]

**7 The turn of the first Millennium CE and different narratives on the reign of Basil II**

Two further laws for the "defence of the poor" were issued by Emperor Basil II (976-1025), whose reign especially due to its military successes in the East and on the Balkans is often seen as apex of medieval Roman power.[107] His early reign between 976 and 989, however, was overshadowed by attempted military coups of two relatives of his predecessors Nikephoros II Phokas and John I Tzimiskes, Bardas Phokas and Bardas Skleros.[108] The succession of two bloody civil wars also contributed to the "apocalyptic" mood of the history of Leo the Deacon, who wrote that "many extraordinary and unusual events have occurred in novel fashion in the course of my lifetime: fearsome sights have appeared in the sky, unbelievable earthquakes have occurred, thunderbolts have struck and torrential rains have poured down, wars have broken out and armies have overrun many parts of the inhabited world, cities and whole regions have moved elsewhere, so that many people believe that life is now undergoing a transformation, and that the expected Second Coming (*deutera katabasis*) of the Savior and God is near, at the very gates."[109]

---

[102] John Scylitzes, *Synopsis*, Nikephoros II, 20, ed. Thurn, pp. 277–78; transl. Wortley, pp. 266–67. Dölger/Müller/Beihammer, *Regesten von 867–1025*, nr 702.
[103] John Scylitzes, *Synopsis*, transl. Wortley, p. 267, n. 78. On the declining popularity of Nikephoros II Phokas and its causes see also Kaldellis, *Streams of Gold*, pp. 51–54 and 63–64.
[104] John Scylitzes, *Synopsis*, John I, 3, ed. Thurn, pp. 286, 48–56; transl. Wortley, pp. 273–74; Leo the Deacon, *History* VI, 8, ed. Hase, p. 103; transl. Talbot/Sullivan, pp. 152–153. Telelis, Μετεωρολογικά φαινόμενα, nr 406.
[105] For parallel reports on weather extremes in Western Europe see Newfield, *The Contours of Disease and Hunger*, p. 484 (nr 304).
[106] Novels of the Macedonian emperors, ed. Svoronos, nr 8–12, pp. 151–84; transl. McGeer, *The Land Legislation of the Macedonian Emperors*, pp. 86–108. See also Sarris, "Large Estates and the Peasantry in Byzantium".
[107] Kaldellis, *Streams of Gold*, p. xxviii.
[108] Kaldellis, *Streams of Gold*, pp. 81–102.

For Basil II, this sequence of portents starts in Leo´s text with a comet in August to October 975 (so still during the reign of John I), which "scholars of astronomy" misinterpreted as sign of future victories, while according to Leo it foretold "bitter revolts, and invasions of foreign peoples, and civil wars, and migrations from cities and the countryside, famines and plagues and terrible earthquakes, indeed almost the total destruction of the Roman empire (…)."[110] Another "sinking" star in August 986 foreboded a defeat of Basil II´s army against the Bulgarians.[111] The sighting of Halley´s Comet between August and September 989, which was equally visible in other parts of Europe[112], was followed by further military defeats and especially a devastating earthquake in Constantinople on 25 October 989, which even damaged Hagia Sophia. Furthermore, so Leo, "harsh famines and plagues, droughts and floods and gales of violent winds (…), and the barrenness of the earth and calamities that occurred, all came to pass after the appearance of the star. But my history will describe these in detail in their place."[113] The reference to drought would find a counterpart in the isotope record from the Sofular cave in Northwestern Asia Minor, which indicates the years around 995 as the driest in the entire 10th century (see **fig. 7**).[114] Tree ring data from modern-day Albania points to very cold conditions in that region in the early 990s.[115] Leo´s history ends, however, shortly after this passage, and the author most probably died at some point before the year 1000. Thus, he did not witness the later military successes of Basil´s II reign, especially his ultimate destruction of the Bulgarian Empire in 1018, which earned him the praise of later historians (since the 12th century as "Bulgar Slayer"), who also wrote under the impression of the severe crisis of the empire emerging under Basil´s successors in the 11[th] century.[116] Therefore, we have few information on the further sequence of "harsh famines and plagues, droughts and floods and gales of violent winds" Leo mentioned as underpinnings of his apocalyptic reading of Basil II´s reign. John Scylitzes for instance informs us that in 1010/1011, "there was a most severe winter; every river and lake was frozen, even the sea itself. And in January of the same year of the indiction a most awesome earthquake occurred; it continued to shake the earth until the ninth of March."[117] Yet, in contrast to the extreme winter of 927/928, Scylitzes does not mention any effects on agriculture or a famine.

---

[109] Leo the Deacon, *History* I, 1, ed. Hase, p. 4; transl. Talbot/Sullivan, pp. 55–56.
[110] Leo the Deacon, *History* X, 6, ed. Hase, p. 169; transl. Talbot/Sullivan, pp. 210–12.
[111] Leo the Deacon, *History* X, 8, ed. Hase, p. 172; transl. Talbot/Sullivan, p. 214; Kaldellis, *Streams of Gold*, pp. 95–96. On this and other observations of this comet see Wozniak, *Naturereignisse im frühen Mittelalter*, pp. 141–42.
[112] Wozniak, *Naturereignisse im frühen Mittelalter*, pp. 106–07.
[113] Leo the Deacon, *History* X, 10, ed. Hase, pp. 175–76; transl. Talbot/Sullivan, pp. 217–18. Kaldellis, *Streams of Gold*, p. 104. On the possible sighting of aurora borealis as one explanation for some of the celestial phenomena described by Leo see Wozniak, *Naturereignisse im frühen Mittelalter*, p. 189.
[114] Fleitmann et al., "Sofular Cave".
[115] PAGES 2k Network consortium, Database.

Leo the Deacon, however, finds a counterpart in the equally apocalyptically inspired 12[th] century Armenian chronicle of Matthew of Edessa:

> "During the reign of Basil, the Greek emperor, and in the year 452 of the Armenian era [1003-1004] a certain star, appearing in the form of fire, arose in the heavens, an omen of the wrath of God towards all living creatures and also a sign of the end of the world. There was a violent earthquake throughout the whole land, to such an extent that many thought that the day of the end of the world had arrived. Like the time of the flood all living creatures shook and trembled, and many fell down and died from fear of the intensity of this wrath. After this outpouring of God´s wrath a plague (…) came upon the area and spreading through many regions, reached Sebasteia [modern-day Siwas, in Byzantine Cappadocia, where many Armenians lived, see **fig. 3**]. This plague clearly manifested itself on men's bodies and, because of its harshness, many had no time to make their confession or take communion. Men and beast diminished from the land, and those remaining quadrupeds roamed about the countryside without anyone to take care of them."[118]

This information on an epidemic in the eastern provinces of the Byzantine Empire is not found in any Greek source, but it fits within a larger pattern of weather extremes and calamities affecting the Middle East in the years around the turn of the first Millennium CE.[119] Again, Matthew of Edessa mentions that "it happened at the beginning of the year 446 of the Armenian era [997-998] that a certain comet arose in the heavens and it became visible with a horrible and dreadful appearance, bright and marvellous."[120] Around the same time, during the reign of the Fatimid Caliph al-Ḥākim (996-1021) in Egypt, "a period of inflation occurred (…). It was caused by an insufficient level of the Nile, which reached only sixteen cubits and a few fingers. Prices rose sharply and wheat was in high demand but was unattainable. The populace lived in a heightened state of fear, women were kidnapped in the streets, and the situation deteriorated. The price of bread reached one dirham for four ratls; then the situation eased when prices dropped."[121] For the area of Baghdad, Bar Hebraeus reports for the winter 998/999 that "severe frost took place (…), and thousands of the palm-trees (…) were destroyed. And those which remained only after very many years acquired straightness".[122] In 1002, first severe cold in March and then an exceptional flood of the Euphrates in April plagued Baghdad before "swarms of locusts appeared in the country of Mosul and in Baghdad and they became very numerous in Shiraz [in Iran]. They left no grass [in the fields] and no leaves on the trees and they even consumed the rolls of cloth which the fullers were bleaching; and of each roll of cloth the fuller was only able to give a rag to its owner. And there was a famine, and a measure of wheat was sold in Baghdad for one hundred and twenty gold dinars. And pillars of fire appeared in the heavens, from the north pole to the middle of the sky."[123]

---

[116] Neville, *Guide to Byzantine Historical Writing*, pp. 124–25; Kaldellis, *Streams of Gold*, pp. 104–05.
[117] John Scylitzes, *Synopsis*, Constantine VIII, 2, ed. Thurn, p. 373, 3–11; transl. Wortley, p. 352. Telelis, *Μετεωρολογικά φαινόμενα*, nr 432.
[118] Matthew of Edessa, *History* I, 46, transl. Dostourian, p. 43.
[119] Some parallel observations exist for Western Europe, see Wozniak, *Naturereignisse im frühen Mittelalter*, pp. 494–96, 617–21.
[120] Matthew of Edessa, *History* I, 41, transl. Dostourian, p. 41.

Between 1005 and 1008, again low Nile floods caused shortages of food and a rise of grain and bread prices in Egypt, which Caliph al-Ḥākim and his officials tried to mitigate with price regulations and drastic measures (such as flogging and public parading) against millers, bakers, hoarders of grain and speculators, who were suspected to take advantage of the misery of the population – obviously with some success, since "prices decreased and harm was averted (…)."[124] Around the same time, in 1007, "snow fell in Baghdad", but the next harvest brought "great abundance" and low prices for wheat. But later, "violent black winds" in the area of Tikrīt to the northwest of Baghdad (see **fig. 3**) "destroyed many houses and tore up very many palm-trees and olive-trees by the roots; and great ships were sunk in the Sea of Persia".[125] In 1010, "(…) swarms of locusts appeared in the country of Mosul, and the nomads raided the country on all sides, and there was also a great pestilence. And the famine waxed strong in the country of Khorasan [in eastern Iran] until one litre of bread was sold for a gold dinar." People would even resort to cannibalism, Bar Hebraeus and other sources tell us.[126] Further reports on floods come from China for the period between 1000 and 1014.[127]

---

[121] al-Maqrīzī, *Ighāthah*, transl. Allouche, p. 31. Telelis, Μετεωρολογικά φαινόμενα, nr 419–420; Hassan, "Extreme Nile floods".
[122] Bar Hebraeus, *Chronography*, transl. Budge, pp. 181–82. Telelis, Μετεωρολογικά φαινόμενα, nr 421; Busse, *Chalif und Großkönig*, p. 388–89.
[123] Bar Hebraeus, *Chronography*, transl. Budge, p. 183. Busse, *Chalif und Großkönig*, p. 389.
[124] al-Maqrīzī, *Ighāthah*, transl. Allouche, pp. 31–33. Telelis, Μετεωρολογικά φαινόμενα, nr 424–427, 430–431; Hassan, "Extreme Nile floods"; Wozniak, *Naturereignisse im frühen Mittelalter*, p. 619.
[125] Bar Hebraeus, *Chronography*, transl. Budge, pp. 183–184. Telelis, Μετεωρολογικά φαινόμενα, nr 428–429; Busse, *Chalif und Großkönig*, p. 389.

Thus, in comparison with Egypt, Mesopotamia, or Persia, despite the apocalyptic expectations of Leo the Deacon, the Byzantine Empire seems to have got of lightly during this cluster of extreme events at the turn of the first Millennium, although we have to take into consideration that the latter historians of the 11th century were less eager to register catastrophes for the reign of Basil II than Leo was. For the late reign of Basil II, we learn from Armenian sources (interpreting this as another sign of divine wrath) that the emperor and his army in 1021/1022 encountered "violent snowstorms" and flooding during a campaign in Armenia (in the area east of Lake Van), which caused severe losses.[128] These extreme conditions could already be attributed to more turbulent climatic period connected with the Oort Solar Minimum (see above).

## 8 The Oort Solar Minimum and a cluster of disasters between 1025 and 1042

As mentioned above, the historians of the late 11th century wrote under the impression of a "decline" from the apex of Byzantine power under Basil II to the almost fatal crisis of the 1070s (especially after the defeat at Manzikert 1071); thus, positive verdicts on the successors of the "Bulgar Slayer" are quite rare, including his brother Constantine VIII.[129] During his reign from 1025 to 1028, so John Scylitzes tells us, "there was a severe drought; even unfailing springs and rivers dried up. The emperor Basil (II) used to spare the poor by not insisting that be paid on time, granting a delay or postponement to those who asked. When he died, two years´ taxes were owing; these payments Constantine demanded immediately, and he also stipulated that taxes be paid on time for the three next years (for that is how long his reign lasted). By demanding five years´ payments in three he ruined not only the poor (*penetes*) and indigent (*aporoi*) but also those who were well off."[130] In the chronicle of John Zonaras (written in the 12th century[131]), the correlation between the stinginess of the emperor, the "ruin" of the poor and the drought is even more explicit: "But since the taxes of two years had not been collected at the death of the Emperor Basil (he granted a deferment of the taxes in order to spare the taxpayers), those of the past and those of the three years of his reign he [Constantine VIII] demanded mercilessly, and that, although all the time of his reign there was drought, so that the poor perished."[132] We may assume that under these conditions, the pressure from above and by nature together with dynamics from below (that is farmers preferring to sell their land to powerful elite neighbours and to work for them as tenants in order to profit from their protection and resources in cases of political unrest, additional tax demands, or – as in the aftermath of 927 – of crop failures and extreme events) worked again for a further shift of the balance of power in the countryside.[133] As Anthony Kaldellis points out, however, the Christian Arab chronicler Yaḥyā of Antioch (11th century) on the contrast praised Constantine VIII "for remitting back taxes and not collecting them on abandoned land".[134]

---

[126] Bar Hebraeus, *Chronography*, transl. Budge, p. 185. Busse, *Chalif und Großkönig*, p. 389. Another plague of locusts occurred in Baghdad in 1018, see Bar Hebraeus, *Chronography*, transl. Budge, p. 185. In general, on the "topos" of cannibalism during famines see Wozniak, *Naturereignisse im frühen Mittelalter*, pp. 731–39.
[127] Zhang, *The River, the Plain, and the State*, p. 37, 110–12; Mostern, *The Yellow River*, pp. 142–44.
[128] Aristakēs Lastiverc῾i, *History*, ed. and transl. Bedrosian, pp. 46–49; Matthew of Edessa, *History* I, 51, transl. Dostourian, p. 47. Telelis, Μετεωρολογικά φαινόμενα, nr 433; Kaldellis, *Streams of Gold*, pp. 131–33.
[129] Kaldellis, *Streams of Gold*, pp. 155–56; Preiser-Kapeller, "Byzantium 1025–1204", p. 60; Krallis, "Historiography as Political Debate".
[130] John Scylitzes, *Synopsis*, Basil II and Constantine VIII, 34, ed. Thurn, pp. 347, 20–348, 1; transl. Wortley, p. 330. Telelis, Μετεωρολογικά φαινόμενα, nr 437. On identifying the "poor" see also Morris, "The Powerful and the Poor".
[131] Neville, *Guide to Byzantine Historical Writing*, pp. 191–96.

The references to a drought during the reign of Constantine VIII, at least, find partial confirmation in similar reports on weather extremes and famines from neighbouring polities. Bar Hebraeus mentions a hailstorm in Baghdad in April 1026, while in winter 1026/1027 "there was an intense cold (…), and the banks of the Euphrates and Tigris were covered with ice, and the palm trees were destroyed. And in Baghdad men used to cross over the small canals on the frozen water, and the farmers were unable to sow seed." Frost periods continued in 1027 and 1028.[135]

In Egypt, under Caliph al-Ḥākim´s successor aẓ-Ẓahir there was an even greater hunger crisis already in the years 1023 to 1025, for which we also have precise information on the price increase (up to ten times the usual value), which the French scholar Thierry Bianquis (1935-2014) evaluated in an important study.[136] The contemporary chronicler al-Muṣābiḥī wrote: "The water (of the Nile) sank incessantly and beyond measure; the lands were not flooded and the ground bore nothing. The people of al-Fusṭāṭ shouted and begged God for moisture. Many people from the city, men and children, went to the mountains with Korans to implore God for water. Bread became scarce in the markets; the masses crowded around the grain."[137]

---

[132] Zonaras, *Epitome* 17, 10, ed. Büttner-Wobst, p. 572–73.
[133] Laiou/Morrisson, *The Byzantine Economy,* p. 106.
[134] Yaḥyā of Antioch, *Chronicle* 15, 71, transl. Pirone, p. 338: "He [Constantine VIII] annulled all claims for sums due from the subjects of the Roman country, what was demanded of them beyond what was due and what was usually collected from the imperial villages that had been destroyed, and from each of their neighbouring communities until they were rebuilt". Kaldellis, *Streams of Gold*, pp. 155–56.
[135] Bar Hebraeus, *Chronography*, transl. Budge, p. 191. Telelis, *Μετεωρολογικά φαινόμενα*, nr 439; Busse, *Chalif und Großkönig*, p. 390.
[136] Bianquis, "Une crise frumentaire dans l'Égypte Fatimide*"*, esp. pp. 100–01. See also Hassan, "Extreme Nile floods".
[137] Cited after Halm, *Die Kalifen von Kairo*, p. 322.

In this time of need, a traditional ceremony turned out to be fatal: in February 1025, on the eve of the Islamic festival of sacrifice, which marks the climax of the pilgrimage season to Mecca, particularly extravagant food arrangements, which the next day were to be served at the caliph's banquet to his exquisite selection of guests, were carried through the capital to present the splendour of the ruler's court. This time, however, such a display made clear to the starving masses the social and material distance between the palace and the rest of the city, and tensions in the population rose. For fear of revolts, not only the broad masses, but also some of the palace guards, including regiments recruited from African slaves, who also could not be fed as usual, were excluded from the banquet. But some of these soldiers entered the palace and according to al-Muṣābiḥī stormed the banquet with the words: "Hunger! We have more right to partake in our Lord's table! (…) Then they attacked the food, fought and stole everything that was prepared: bread, roast meat and sweets; they stole bowls, trays and bowls. Bad thing! They took 300 plates with them, and those present could hardly believe that they had escaped and got away with intact skin."[138] Cairo and al-Fusṭāṭ experienced a multiple breakdown in the symbolic communication between rulers, court society, army and people in these days. Acceptance for the regime dwindled to a dangerous extent.[139] Relief finally brought sufficient Nile floods, which allowed a return to the usual generosity of the caliph, especially towards his troops.[140]

In Byzantium, the drought of the reign of Constantine VIII, however, reportedly ended after his death with the coronation of Emperor Romanos III Argyros (1028-1034), husband of Constantine´s daughter Zoë; Skylitzes wrote: "In those days God caused an adequate amount of rain to fall and the crops were abundant, especially the olives."[141] Soon, however, celestial phenomena foretold havoc, and weather extremes and hardship followed: "In October [1029], the thirty-first of that month, the fall of a star occurred, following a path from west to east, and on that day the Roman army suffered a severe defeat in Syria (…). And rain fell in torrents continuously until the month of March. The rivers overflowed and hollows turned into lakes, with the result that nearly all the livestock was drowned, and the crops were levelled. This was the cause of a severe famine in the following year."[142]

The sequence of calamities continued two years later:

---

[138] Cited after Halm, *Die Kalifen von Kairo*, p. 322.
[139] Oesterle, *Kalifat und Königtum*, pp. 160–62. Cf. also Halm, *Die Kalifen von Kairo*, pp. 319–24; Brett, *The Fatimid Empire*, pp. 163–64; Chipman/Avni/Ellenblum, "Collapse, affluence, and collapse", 202–03.
[140] On the role of the military in Fatimid Egypt see Sanders, "The Fāṭimid State, 969–1171", pp. 154–57.
[141] John Scylitzes, *Synopsis*, Romanos III Argyros, 2, ed. Thurn, p. 376, 77–78; transl. Wortley, p. 355. Telelis, *Μετεωρολογικά φαινόμενα*, nr 440; Olson, *Environment and Society in Byzantium*, p. 183.
[142] John Scylitzes, *Synopsis*, Romanos III Argyros, 3, ed. Thurn, p. 377, 4–12; transl. Wortley, p. 356.

> "On Friday, 28 July, at the second hour of the night, a star fell from south to north, lighting up the whole earth, and shortly afterwards there were reports of disasters afflicting the Roman empire (…). This year [1032] famine and pestilence afflicted Cappadocia, Paphlagonia, the Armeniakon theme and the Honoriad, so grave that the very inhabitants of the themes abandoned their ancestral homes in search of somewhere to live. The emperor [Romanos III Argyros] met them on his return to the capital from Mesanakata [in central Asia Minor, to which a campaign had let him[143]] and, unaware of the reason for this migration, obliged them to return home, providing them with money and the other necessities of life. And Michael, who was then governing the church of Ankyra, performed virtuous works, sparing nothing which might procure the survival of the victims of famine and pestilence. On Sunday, 13 August, at the first hour of the night, AM 6540, there was a severe earthquake. The emperor came into the capital and Helena, his former wife, having died, he distributed many alms on her behalf. In that year on 20 February [1033] a star traversed from north to south with noise and commotion. It was visible until 15 March, and there was a bow above it. On 6 March, third hour, there was an earthquake."[144]

Yet beyond the accumulation of portents and catastrophes, John Scylitzes uses these calamities to hint at the efforts of Romanos III (and a member of the church elite) to mitigate the need of those afflicted by famine, disease, and earthquakes.[145] This is also true for the last period of the emperor´s reign:

> "(…) On the seventeenth of February [1034], there was an earthquake and the cities of Syria suffered severely. (…) For some time, the eastern themes had been consumed by locusts, compelling the inhabitants to sell their children and move into Thrace. The emperor gave to every one of them three pieces of gold and arranged for them to return home. The locusts were finally carried away by a powerful wind, fell into the high sea off the Hellespont and perished. They were washed up onto the shore where they covered the sand of the beach. The emperor renovated the aqueducts which bring water into the city and also the cisterns which receive that water. He restored the leper house and every other hospice which had been damaged by the earthquake. In a word, every good work was his concern. But he was afflicted by a chronic disease; his beard and his

---

[143] Kaldellis, *Streams of Gold*, p. 163.
[144] John Scylitzes, Synopsis, Romanos III Argyros, 10–12, ed. Thurn, pp. 385, 52–386, 81; transl. Wortley, pp. 364–365. See also Aristakēs Lastivercʿi, *History*, ed. and transl. Bedrosian, pp. 76–77 (dating the portent to 1033, but to the reign of Michael IV). Wozniak, *Naturereignisse im frühen Mittelalter*, p. 621; Dagron, "Quand la terre tremble".
[145] Kaldellis, *Streams of Gold*, pp. 163–64.

> hair fell out. It was said he had been poisoned by John, who later became *orphanotrophos*."[146]

The "good works" of Romanos III are contrasted with the intrigues of John, a eunuch at the court, who arranged for a love affair between his brother Michael and empress Zoë. Thus, the empress reportedly supported a plot to get rid of her husband Romanos III by poisoning and murder, which allowed her to marry Michael [IV], who became the new ruler.[147] John, who held the office of *orphanotrophos*[148], however, was the true power behind the throne, especially also since his brother suffered from severe illness (the description in the sources hints at epilepsy)[149], which John Skylitzes together with portents and calamities interpreted as signs of divine distaste for the new regime:

> "But it was clearly shown from the outset that what had transpired was not pleasing to God. At the eleventh hour of Easter Day [14 April 1034] there was an unendurable hailstorm, so violent that not only the trees (fruit-bearing and otherwise) were broken down, but also houses and churches collapsed. Crops and vines were laid flat to the ground; hence there ensued a great shortage of all kinds of produce at that time. There was a falling star about the third hour of the night on the Sunday after Easter; the brilliance of its shining put all the stars into the shade and, for many, it looked like the rising sun. And the emperor [Michael IV] became possessed of a demon; those close to him, using fine phrases, called it a madness-causing disease, but it endured to the end of his life. He received no relief either by divine might or from doctors but was grievously tormented and tortured."[150]

To make his message clear, Skylitzes connects the renewed outbreak of the plague of locusts with a vision of a eunuch (meaning John, the Orphanotrophos, of course) who carries three sacks full of vermin:

> "The swarms of locusts which had expired (as we reported) on the sands of the shore of the Hellespont now spontaneously regenerated and overran the coastal regions of the Hellespont again, devastating the Thrakesion theme for three whole years. Then they appeared in Pergamon but perished there, as one of the bishop's servants saw beforehand

---

[146] John Scylitzes, *Synopsis*, Romanos III Argyros, 17, ed. Thurn, p. 389, 54–69; transl. Wortley, pp. 367–68. Cf. also Zonaras, *Epitome* 17, 12, ed. Büttner-Wobst, pp. 580–81. Wozniak, *Naturereignisse im frühen Mittelalter*, p. 576.
[147] Preiser-Kapeller, "Byzantium 1025–1204", p. 61; Kaldellis, *Streams of Gold*, pp. 164–65. The rumours about the plot of Zoë also made it to Armenian sources, see Aristakēs Lastiverc'i, *History*, ed. and transl. Bedrosian, pp. 66–69.
[148] For this office, see Kazhdan, "Orphanotrophos".
[149] Kaldellis, *Streams of Gold*, pp. 165–66.
[150] John Scylitzes, *Synopsis*, Michael IV, 2, ed. Thurn, p. 393, 45–57; transl. Wortley, p. 371

> in a vision (not a dream, for he was awake). It was as though he saw a eunuch dressed in white, of radiant appearance. [This apparition] was ordered to open and empty the first of three sacks lying before him, then the second and, after that, the third. He did as he was commanded; the first sack poured out snakes, vipers and scorpions; the second, toads, asps, basiliscs, horned serpents and other venomous creatures; the third, beetles, gnats, hornets and other creatures with a sting in the tail. The man stood there speechless; the bright apparition stood close to him and said: 'These came and will come upon you because of your transgression of God's commandments and the desecration of the emperor Romanos which has taken place and the violation of his marriage bed.' That is what happened so far."[151]

God also rejects any appeal for a relief from another climatic calamity in 1036: "Because there was a drought and for six whole months no rain had fallen, the emperor's brothers held a procession, John carrying the holy *mandylion*, the Great Domestic the Letter of Christ to Abgar, the *protobestiarios* George the holy Swaddling Bands. They travelled on foot from the Great Palace to the church of the exceedingly holy Mother of God at Blachernae. The patriarch and the clergy made another procession, and not only did it not rain but a massive hailstorm was unleashed which broke down trees and shattered the roof tiles of the city. The city was in the grip of famine so John purchased one hundred thousand bushels of grain in the Peloponnese and in Hellas; with this the citizens were relieved."[152] With the last sentence, however, Skylitzes has to admit that some of the relief measures of his villain, John Orphanotrophos, proved effective – and we also learn that the drought affecting the environs of Constantinople did not prevent a surplus of harvest in central and southern Greece to be purchased for the capital.

In another episode reported by Scylitzes, also the actions of Emperor Michael IV, who spent a long time of his reign in Thessalonike, where he hoped for a healing from his illness at the intervention of Saint Demetrios, are quite efficient for the mitigation of a supply shortfall:

> "In AM 6546, sixth year of the indiction, there was an earthquake on 2 November [1037] about the tenth hour of the day, and the earth continued to tremble into and throughout the month of January [1038]. There was a famine in Thrace, Macedonia, Strymon and Thessalonike, right into Thessaly. When the clergy of Thessalonike accused Theophanes the Metropolitan of withholding their customary allowance, the emperor (who was staying there) tried admonishing him, urging him not to deprive the personnel of the church of the allowance stipulated in the law. When the bishop showed himself recalcitrant and inflexible, the emperor realised he would have to circumvent him with a subterfuge to punish his avarice. He therefore sent one of his servants to him requesting the loan of one kentenarion [100 pounds] until gold was delivered from Byzantium. The bishop denied with oaths that he had any more than thirty pounds on hand but the emperor did not let this stand in his way. He sent and scrutinised the man's treasury and found thirty-three kentenaria of gold. Out of this he paid the clergy what was owing to them since the first year of Theophanes' episcopate until the present hour: the rest he distributed to the poor. He expelled the metropolitan from the church and restricted him to an estate. (…)."[153]

---

[151] John Scylitzes, *Synopsis*, Michael IV, 4, ed. Thurn, pp. 394, 77–395, 94; transl. Wortley, p. 372. Krallis, "Historiography as Political Debate".

[152] John Scylitzes, *Synopsis*, Michael IV, 10, ed. Thurn, p. 400, 39–49; transl. Wortley, pp. 377–78.

Otherwise, however, for Scylitzes the regime around Michael IV and John Orphanotrophos could not hope for divine forgiveness, and thus the sequence of calamities affecting them, and the empire continued:

"The emperor was still afflicted by the demon and, finding no relief, he sent two pieces of gold for each priest in all the themes and the islands, one for each monk. He also stood godfather at the baptism of new-born children, giving each one a single piece and four miliarisia, but none of this did him any good. In fact, the condition worsened and in addition he was afflicted by dropsy. That year there were continuous earthquakes and frequent heavy rainfalls while, in some of the themes there was such an epidemic of quinzy [*kynagches nosima*, maybe diphtheria] that the living were unable to carry away the dead. On 2 February, eighth year of the indiction, AM 6548 [1040], there was an appalling earthquake; other places and cities suffered too. Smyrna was a pathetic sight for its most beautiful buildings fell down and many of the inhabitants lost their lives."[154]

John Skylitzes is even more explicit in his "moral meteorology" in another passage:

"Most of the time the emperor Michael resided at Thessalonike where he frequented the tomb of the wondrously victorious martyr Demetrios in the sincere hope of finding relief from his illness. He had nothing whatsoever to do with affairs of state other than those which were absolutely necessary; the administration and the handling of public business rested entirely on John's shoulders and there was no imaginable form of impurity or criminality that he did not search out for the affliction and mistreatment of the subjects. It would be a Herculean task to list them all. Everybody living under this grievous tyranny persisted in interceding with the Deity, appealing for some relief. God frequently shook the earth; the inhabited world was assailed by awesome and fearful [portents]: comets appearing in the sky, storms of wind and rain in the air, eruptions and tremblings on earth. In my opinion, these things presaged the forthcoming unparalleled catastrophe for the tyrants."[155]

---
[153] John Scylitzes, *Synopsis*, Michael IV, 13, ed. Thurn, p. 402, 81–5; transl. Wortley, p. 379.
[154] John Scylitzes, *Synopsis*, Michael IV, 18–19, ed. Thurn, p. 405, 67–79; transl. Wortley, p. 381.

Before the final act towards the ultimate punishments of "tyrants" can take place, further calamities affect the empire and the capital: "There was a drought that year [1040], so severe that copious springs and ever-flowing rivers almost dried up. There was a fire at the Arsenal [in Constantinople] on 6 August and all the ships that were moored there were burnt together with their fittings."[156] On 10 December 1041, Michael IV died.[157] John the Orphanotrophos, however, had taken precautions for this by introducing his nephew Michael to the court. He found the approval of Empress Zoë, who adopted him. Thus, Michael V accessed the throne on the day after the death of his uncle.[158] Skylitzes, however, illustrates the shaky basis of this regime from the beginning: "In the very same hour at which he received the diadem Michael [V] was afflicted with vertigo and swimming in the head. He almost fell over; they were only just able to revive him with sweet oils, perfumes and other aromatic substances. the earth was a-tremble throughout the four months of his reign."[159] As a matter of fact, Michael V causes his own downfall within a few months by first exiling his uncle John, the actual mastermind of the rise of his family, to a monastery, and second by an attempt to do the same with his adoptive mother empress Zoë. This in April 1042 led to a general rebellion in Constantinople, where large groups of the population were still loyal to the Macedonian dynasty of which Zoë and her sister Theodora were the last scions.[160] Michael V was toppled from the throne and eventually blinded together with his uncle John; Scylitzes finished his description of the "unparalleled catastrophe for the tyrants" with the sentence: "Michael fervently entreated that his uncle [John Orphanotrophos] be blinded before him because he was the cause and instigator of all the evils that had taken place and that is what happened."[161]

If read in continuation and in context (and not as single pieces of meteorological data), the narrative function of the descriptions of extreme events and calamities provided by John Skylitzes becomes clear. In contrast and beyond Skylitzes´ interpretation, it is hard to estimate to what extent these portents and catastrophes undermined the legitimation of and support for the regime of John Orphanotrophos and his brother and nephew, especially since Skylitzes himself in two cases has to concede the efficiency of relief measures of John respectively Michael IV.[162]

---

[155] John Scylitzes, *Synopsis*, Michael IV, 21, ed. Thurn, p. 408, 51–63; transl. Wortley, pp. 383–84. See also Kaldellis, *Streams of Gold*, p. 168. The long stays of Michael IV in Thessaloniki are also mentioned in Aristakēs Lastiverc῾i, *History*, ed. and transl. Bedrosian, pp. 72–73.
[156] John Scylitzes, *Synopsis*, Michael IV, 24, ed. Thurn, p. 411, 46–50; transl. Wortley, p. 386.
[157] John Scylitzes, *Synopsis*, Michael IV, 29, ed. Thurn, p. 415, 50–56; transl. Wortley, p. 390.
[158] Kaldellis, *Streams of Gold*, pp. 174–75; Preiser-Kapeller, "Byzantium 1025–1204", p. 61.
[159] John Scylitzes, *Synopsis*, Michael V, 1, ed. Thurn, p. 417, 79–83; transl. Wortley, p. 392. Cf. also Michael Attaleiates, *History* 4, 9, ed. and transl. Kaldellis/Krallis, pp. 26–29.
[160] Kaldellis, *Streams of Gold*, pp. 175–78.
[161] John Scylitzes, *Synopsis*, Michael V, 2, ed. Thurn, p. 420, 96–421, 2; transl. Wortley, pp. 395–96.

The clustering of meteorological extremes and calamities in the 1030s as such finds again confirmation in similar descriptions for neighbouring regions. The famine affecting the Byzantine provinces in Asia Minor in 1032/1033 also raged further to the east in Armenia and Northern Mesopotamia, as Matthew of Edessa indicates: "At the beginning of the year 481 of the era of the Armenian calendar [1032-1033] there was a severe famine throughout the entire land. Many people died because of this famine, and many sold their women and children for want of bread. Because of the intensity of the hardships, whenever one spoke, he yielded up his soul. In this manner the land was consumed by famine."[163]

For Baghdad and wider regions of the Islamic world, catastrophe was equally foretold by a portent:

> "And in the year four hundred and twenty-three of the Arabs (1031 CE) a woman in Baghdad gave birth (to a being) which was like an ill-formed serpent. He had the head of a man, and a mouth and a neck, and he was without hands and without feet. And, moreover, when he fell upon the ground, he spoke and said, 'Four years from now a famine shall make an end of the children of men, unless men, and women, and children, and the beasts go forth and weep before the Lord, so that He may make His rain to descend'. And when the Caliph heard this, he commanded that all the people should go outside [the city] and make supplication. And because many did not believe this report, very few went out. And in that year [1031-1032] the water froze in Bagdad and the red sand descended as rain, and the trees were destroyed and produced no fruit at all the season. And there was so great a famine in the wilderness that the nomads who lived there ate their camels and their horses, and even their children. (…) And they were in tribulation not only because of the famine (or, want of food), but also through thirst which was due to the scarcity of water, and they came and camped by the rivers (or, canals) which were in the neighbourhood of the towns and villages. And there was a pestilence in India and in all Persia; forty thousand biers with dead men on them were taken out from Isfahan in one week. And in Baghdad also there was not a single house left in which there was not wailing. And in Mosul four thousand young men died of the disease of inflammation of the eye-lids."[164]

---

[162] Kaldellis, *Streams of Gold*, p. 168, on other sources indicating that Michael IV was relatively popular. See also Krallis, "Historiography as Political Debate".
[163] Matthew of Edessa, *History* I, 60, transl. Dostourian, p. 55. Telelis, Μετεωρολογικά φαινόμενα, nr 443.

A disease maybe akin to the one reported for Byzantium for 1039/1040 (maybe diphtheria) raged in Baghdad a few years earlier in 1033/1034, when heavy storms and rainfalls also affected Northern and Southern Mesopotamia (in general, the Kuna Ba isotope date from Northern Iraq indicates much more humid conditions on average when compared with the 10th century, see **fig. 6**).[165] For 1037, Bar Hebraeus reports again snowfall in Baghdad, followed by an intense cold. Similar phenomena occurred in January 1039, followed by unusual heat in March and another frost period a few days later.[166] In 1042, hailstorms destroyed more than 30 villages around Baghdad.[167] For the years between 1042 and 1045, we read about famine in various regions of Western Europe.[168] Equally, in China the 1030s and 1040s were plagued by various catastrophes such as droughts, floods, famine, earthquakes and epidemics among humans and animals, culminating in a extremely devastating flood of the Yellow River in 1048, followed by another famine and large movements of refugees.[169] Some of these global climate fluctuations, in addition to or as further effect of the impact of the Oort Solar Minimum, may have been connected to strong La Niña events identified also on the basis of proxies from South Asia and the Americas.[170]

## 9 Further climatic anomalies and the migrations from the Steppes

Scylitzes continued his focus on portents to the reign of Constantine IX Monomachos (June 1042-January 1055), the last husband of Empress Zoë, who died in June 1050[171]: "Those were the first deeds of (Constantine IX) Monomachos in the tenth year of the indiction. On the eleventh of October, eleventh year of the indiction, AM 6551 [1042], a comet appeared travelling from east to west and it was seen shining during the whole month; it presaged the forthcoming universal disasters."[172] We do not find, however, a similar density of meteorological phenomena and calamities as in the narrative of Skylitzes for the years 1025 to 1042. For September 1043, a year, when Constantine IX also had to face a military rebellion and a surprise attack of the Rus on Constantinople, Skylitzes informs us, "a wind blew so violently that almost the entire fruit of the vine was destroyed."[173] A major disaster of meteorological and epidemiological character in Skylitzes "Synopsis" only hit Constantinople again towards the end of Constantine IX´s reign in 1054: "In the seventh and eighth years of the indiction the capital was visited by a plague; the living were unequal to the task of bearing away the dead. In the summer of the seventh year of the indiction [July-September 1054] there was a great hailstorm which caused many deaths, not only of animals but of men too. The emperor had an attack of gout, a familiar affliction for him, and was lying in the Mangana monastery which he had recently built. A further illness followed on the first one and he was near to death; the question of whom they should establish on the imperial throne was debated by those who held the highest positions in the palace."[174]

---

[164] Bar Hebraeus, *Chronography*, transl. Budge, pp. 193–94. Telelis, Μετεωρολογικά φαινόμενα, nr 443, 444. Busse, *Chalif und Großkönig*, pp. 390–91.
[165] Bar Hebraeus, *Chronography*, transl. Budge, p. 194. Telelis, Μετεωρολογικά φαινόμενα, nr 445–446, 449–450. For the Kuna Ba data see Sinha et al., "Role of climate".
[166] Bar Hebraeus, *Chronography*, transl. Budge, p. 199. Telelis, Μετεωρολογικά φαινόμενα, nr 456; Busse, *Chalif und Großkönig*, p. 391.
[167] Bar Hebraeus, *Chronography*, transl. Budge, p. 200. Telelis, Μετεωρολογικά φαινόμενα, nr 461.
[168] Wozniak, *Naturereignisse im frühen Mittelalter*, pp. 622–23.
[169] Zhang, *The River, the Plain, and the State*, pp. 1–3, 100–02, 224–26; Mostern, *The Yellow River*, pp. 155–60, 166–68.
[170] Campbell, *The Great Transition*, pp. 39–43.
[171] Kaldellis, *Streams of Gold*, pp. 180–81, 201; Preiser-Kapeller, "Byzantium 1025–1204", pp. 61–62, 64.
[172] John Scylitzes, *Synopsis*, Constantine IX, 2, ed. Thurn, p. 423, 56–424, 62; transl. Wortley, p. 399.

Around the same time, extreme cold affected Armenia in 1054 and 1055[175], while low Nile floods causes a "dearth of grain" in Egypt between 1052 and 1055, aggravated by speculations of members of the elite and even the Fatimid Caliph al-Mustanṣir (1036-1094) himself.[176] The Fatimid government also asked for help from Constantine XI Monomachos. The emperor after April 1054 agreed to ship 400,000 *irdabb*, which are 2,700 tons of grain, to Egypt.[177] Yet the death of Constantine IX in January 1055 dashed all hopes of the Fatimid regime for support from Constantinople, since Empress Theodora, the sister of Zoë and last scion of the Macedonian dynasty, decided to cancel the deal.[178]

The dynamic vizir al-Yāzūrī, however, was able to convince Caliph al-Mustanṣir to end speculations in grain and in general to confine price rigging. As al-Maqrīzī indicates, "al-Yāzūrī performed his task superbly for twenty months until the harvest of the crops two years later, which relieved the populace and ended inflation. People did not suffer in the least, thanks to his good administration. After the vizir [al-Yāzūrī] was killed [in March 1058], the state enjoyed neither righteousness nor stability. The affairs of the state were in disarray, and no praiseworthy or efficient vizir was appointed. The office of vizir became highly discredited". This violent change of administration proved fatal in the next, even more severe supply crisis in Fatimid Egypt in the 1060s (see below).[179]

---

[173] John Scylitzes, *Synopsis*, Constantine IX, 7, ed. Thurn, p. 433, 38–39; transl. Wortley, p. 407. Telelis, *Μετεωρολογικά φαινόμενα*, nr 462; Kaldellis, *Streams of Gold*, pp. 184–87.
[174] John Scylitzes, *Synopsis*, Constantine IX, 7, ed. Thurn, p. 477, 74–82; transl. Wortley, p. 445. Telelis, *Μετεωρολογικά φαινόμενα*, nr 472. On the death of Constantine IX see also Michael Psellos, *Chronographia* VI, 201–203, ed. Reinsch, pp. 195–96.
[175] Telelis, *Μετεωρολογικά φαινόμενα*, nr 471 and 474.
[176] al-Maqrīzī, *Ighāthah*, transl. Allouche, pp. 33–36. Telelis, *Μετεωρολογικά φαινόμενα*, nr 469, 475; Hassan, "Extreme Nile floods".
[177] Dölger/Wirth, *Regesten von 1025–1204*, nr 912; Miotto, *Ο ανταγωνισμός Βυζαντίου και Χαλιφάτου των Φατιμιδών*, pp. 251–52 (with reference to the Arabic sources); Halm, *Die Kalifen von Kairo*, pp. 381–82; Felix, *Byzanz und die islamische Welt*, pp. 119–20. Cf. in general Jacoby, "Byzantine Trade with Egypt", and for the imperial granaries Cheynet, "Un aspect du ravitaillement de Constantinople".
[178] Dölger/Wirth, *Regesten von 1025–1204*, nr 929a.

One cause for the decision of Empress Theodora to cancel the delivery of grain to Egypt was the advance of the Seljuks from Iran, who in 1055/1056 took over control in Baghdad and replaced the Buyids as "protectors" and de facto rulers in Mesopotamia.[180] The upheavals connected to warfare and regime change between 1055 and 1060 were aggravated by plagues of insects and diseases, so that famine raged in Iraq, Syria, and Persia, while sickness affected Baghdad, where "great swarms of flies which polluted the air, and more than one third of the population perished." Bar Hebraeus even claims the within three month 1.65 million people died in Bukhara and within two months 236,000 in Samarkand (both unrealistic numbers); and "it is said that from the beginning of the world never there was such a plague as this."[181] While first raids of Seljuk and other Turkmen units perturbed Armenia and eastern Anatolia, the area, largely under Byzantine rule, according to Matthew of Edessa in 1058 was plagued by an long severe winter, followed by drought, harvest failure and famine; "on the other hand, at the beginning of the next year there was plenteousness and abundance of all types of foodstuffs (…)."[182] This abundance, however, was soon killed off by the increasing amount of Seljuk and Turkmen advances towards Armenia, Anatolia, and Syria; in 1064 they even conquered the former royal Armenian capital of Ani (see **fig. 3**), since 1045 under Byzantine administration.[183]

One cornerstone of Ellenblum´s scenario was the notion of climate-induced stress on the steppe nomads in Central Asia and resulting migration of Turkmen groups under the leadership of the Seljuk dynasty towards the south into Transoxania and Eastern Iran (Khorasan) in the 1040s. Ellenblum partly based his study on an earlier analysis of Richard Bulliet, who attributed the Seljuk invasion and the "decline" of the wealth of Eastern Iran (including the cotton "industry" which had expanded the centuries before) to a severe cold period (the "Great Chill") in the 11$^{th}$ century.[184] As a matter of fact, some natural scientific proxy data sets (one also cited by Bulliet) indicate severely arid and/or cold conditions during parts of the late 10$^{th}$ or early 11$^{th}$ century; they are, however, located at considerable distance from the areas of settlement of the Turkmen before their migrations, respectively is there no overlap between the periods of the adverse climate conditions and the movements of the Seljuks to the south.[185]

---

[179] al-Maqrīzī, *Ighāthah*, transl. Allouche, pp. 35–36. Telelis, Μετεωρολογικά φαινόμενα, nr 475.
[180] Kaldellis, *Streams of Gold*, pp. 197–99, 215.
[181] Bar Hebraeus, *Chronography*, transl. Budge, p. 209. Telelis, Μετεωρολογικά φαινόμενα, nr 477; Busse, *Chalif und Großkönig*, pp. 391–92.
[182] Matthew of Edessa, *History* II, 11, transl. Dostourian, p. 94. Telelis, Μετεωρολογικά φαινόμενα, nr 478, 486 (with a wrong dating to the year 1068–1069).
[183] Aristakēs Lastiverc'i, *History*, ed. and transl. Bedrosian, pp. 308–13; Beihammer, *Byzantium and the Emergence of Muslim-Turkish Anatolia*, pp. 27, 55, 111–15; Kaldellis, *Streams of Gold*, pp. 234–35; Preiser-Kapeller, "Byzantium 1025–1204", p. 63.
[184] Bulliet, *Cotton, Climate, and Camels*. See also Koder, "Zeitenwenden".

Furthermore, based on a systematic and detailed survey of the written sources and partly also of proxy data, three scholars recently and independently rejected the Bulliet-Ellenblum scenario; Jürgen Paul in 2016 wrote: "In all, the article rejects Bulliet's causation chain and proposes that the Ghuzz-Seljuq migration into Transoxiana was due to political reasons rather than induced by climate change."[186] Deborah Tor in 2018 stated that "this catastrophe was not due to 'climate, cotton and camels' [the title of Bulliet´s book] – in fact, Khurāsān was doing very well until the 1150s – but to concrete human agency and action."[187] And most recently, Yoshua Frenkel (2019) confirmed that "that the sources do not provide decisive evidence to support a meteorological interpretation as the prime explanation of the massive human movement across the Steppes/Iranian frontier during the eleventh century".[188] Thus, although we find some references to climatic extremes and epidemics in Eastern Iran and Central Asia from Bar Hebraeus around the time of the Seljuk advance (see above), for instance, specialists on the region do not find sufficient evidence in the sources for a climate-induced migration and collapse-scenario as proposed by Bulliet and Ellenblum, who otherwise argued for a primate of historiography over proxy data.

The Byzantines, however, faced nomadic invaders from the Steppes not only in Anatolia, but equally at the Danube, as also prominently discussed in Ellenblum´s chapter on Byzantium. The Pechenegs had controlled the steppes north of the Black Sea since the late 9th century, but since the 1030s were increasingly under pressure from new steppe formations in the East (the Oghuz and the Cumans) and the Russian princedoms in the north. Eventually, some groups crossed the Danube and after some devastating warfare came to an arrangement with Constantinople, which included the establishment of a semi-independent "Patzinakia" in modern-day Bulgaria, from where, however, raids to other provinces of empire and even involvement in Byzantine civil wars were undertaken. Only Alexios I Komnenos was able to ultimately defeat them and to avoid the emergence of an enduring Pecheneg polity in 1091 CE.[189]

At this "front" we also find some references to extremely cold conditions, as in 1048, when "it was toward the end of autumn and winter about to begin, the sun being in Capricorn, when a very strong wind arose from the north so that the river [Danube] froze to a depth of fifteen cubits. All guard duties being relaxed, Tyrach [one leader of the Pechenegs] seized the opportunity for which he prayed: he crossed the Danube with all the Patzinaks [Pechenegs], eighty thousand in number they say. They installed themselves on the other side, razing and devastating everything they came across."[190] We hear, however, also about disease among the newcomers from the Steppes: "The enemy [the Pechenegs] were not yet used to these lands that were foreign to them and were afflicted by a pestilent disease; they were, moreover, not used to fighting against the Roman phalanxes. So they did not even attempt to raise up arms against them, but gave up as hostages their own rulers and commanders and thus pretended to have been subdued, obtaining a reprieve in this way."[191]

---

[185] See also Campbell, *The Great Transition*, pp. 48–49.
[186] Paul, "Nomads and Bukhara".
[187] Tor, "The Eclipse of Khurāsān".
[188] Frenkel, "The Coming of the Barbarians".
[189] Curta, "The Image and Archaeology of the Pechenegs"; Meško, "Pecheneg groups in the Balkans".

Yet while the Pechenegs were able to establish a durable presence in the Balkans[192], weather and disease together with enemy attacks brought the almost complete destruction to another group trying to cross the Danube in 1064/1065, the Oghuz (*Ouzoi* in Greek sources)[193]; Michael Attaleiates wrote:

> "Such was the state of his [Emperor Constantine X Dukas, r. 1059-1067] preparation [for a campaign against the Oghuz] when messengers from those who had been dispatched to the Danube returned post-haste with the news that that nation had been utterly destroyed. For the captured generals had managed to escape their captivity and explained how this destruction had occurred, saying that the leaders of the Ouzoi had, at the instigation of the Roman authorities in the cities along the Danubian shores, embarked on ships and crossed the Danube, returning to their own lands. Among those who were left behind, however, a vast horde still, some were devastated by an epidemic disease and hunger and were only half alive, while others had been defeated by the Bulgarians and the Pechenegs who were in proximity and were utterly annihilated by iron and the hooves of horses and were even crushed by their own wagons. And so they were killed contrary to all human expectation and those who at one point thought that they would prevail over all others were now held in little regard. In fact, the reports were not far from the truth. (…) As for this Scythian nation, some crossed the Danube and were destroyed by a famine against which there was no recourse, for they had no food and no hope of foraging for it, as their land had neither been ploughed nor sown."[194]

---

[190] John Scylitzes, *Synopsis*, Constantine IX, 7, ed. Thurn, p. 458, 40–46; transl. Wortley, p. 429. Telelis, *Μετεωρολογικά φαινόμενα*, nr 467; Wozniak, *Naturereignisse im frühen Mittelalter*, p. 499; Kaldellis, *Streams of Gold*, pp. 192–94; Dölger/Wirth, *Regesten von 1025–1204*, nr 879c. Cold conditions between 1046 and 1048 are also indicated in tree ring data from Albania, see PAGES 2k Network consortium, Database.
[191] Michael Attaleiates, *History* 7, 2, ed. and transl. Kaldellis/Krallis, pp. 52–53.
[192] Kaldellis, *Streams of Gold*, pp. 199–201.
[193] Kaldellis, *Streams of Gold*, p. 236.

Similar descriptions can be found in the continuation of Scylitzes´ chronicle and the work of Zonaras.[195]

It is hard to tell if the factors described as bringing about the defeat of the Oghuz also originally had motivated their decision to attempt a migration/invasion across the Danube into Byzantine territories, thus confirming Ellenblum´s scenario at least for this region. A prevalence of rather unstable and extreme climatic conditions across parts of Southwestern Asia and the Eastern Mediterranean during the 1060s is again confirmed by parallel descriptions for Iraq and Egypt. In Baghdad, the month October to December 1063, for instance, were characterized by "fierce heat", accompanied by "sickness and pestilence", while January 1064 brought "intense cold", with the Tigris frozen and snowfall.[196] Furthermore, severe floods plagues China in the 1060s, and droughts rages between 1068 and 1070; these phenomena hint a larger-scale climate anomaly across Afro-Eurasia, as mentioned above maybe to be connected with strong El Niño events as also reconstructed from South American proxy data. It also affected precipitation conditions in East Africa in the headwaters of the Nile.[197]

**10 The Great Calamity in Fatimid Egypt and the Byzantine loss of Anatolia**

Due to this precipitation shortfalls, Fatimid Egypt in the 1060s and early 1070s was plagued by a sequence of low Nile floods, aggravated by the instability of the regime after the death of Vizir al-Yāzūrī in March 1058 and its inability to take efficient measures to mitigate the catastrophic supply shortfalls, not even for the core guard regiments of the Caliph. This contributed to the outbreak of civil war between competing factions of the army, which finally would even plunder the Caliphal palaces in Cairo, and an almost-collapse of Fatimid reign in Egypt. Al-Maqrīzī sums up: "During the reign of [Caliph] al-Mustanṣir occurred the famine that had an atrocious effect and left a horrid memory. It lasted seven years and was caused by the weakness of the [Caliph´s] authority, the deterioration of the affairs of state, the usurpation of power by the military commanders, the continuous strife among the Bedouins, the failure of the Nile to reach its plenitude, and the absence of cultivation of the lands that had been irrigated. This began in 457/1064-65. It resulted in rising prices and increased famine and was followed by an epidemic. The lands remained uncultivated, and fear prevailed. Land and sea routes became unsafe, and travel became impossible without a large escort; otherwise, one would be exposed to danger."[198]

---

[194] Michael Attaleiates, *History* 14, 9 and 12, ed. and transl. Kaldellis/Krallis, pp. 154–57 and 158–59.
[195] John Scylitzes continuatus, *Chronicle*, ed. and transl. McGeer, pp. 64–65. See also Zonaras, *Epitome* 18, 9, ed. Büttner-Wobst, p. 680, and Wozniak, *Naturereignisse im frühen Mittelalter*, p. 658.
[196] Bar Hebraeus, *Chronography*, transl. Budge, p. 216. Telelis, Μετεωρολογικά φαινόμενα, nr 481–482.
[197] Hassan, "Extreme Nile floods"; Zaroug/Eltahir/Giorgi, "Droughts and floods"; Zhang, *The River, the Plain, and the State*, pp. 194–95; Mostern, *The Yellow River*, pp. 156–57, 168–70.

As the sequence of factors for the later so-called "Great Calamity" mentioned by al-Maqrīzī makes clear, he lists the shortfall of the Nile floods as one cause among others, especially political and institutional ones. As we have read for earlier low Nile floods of the 11th century, the Fatimid governments then had been able to mitigate the distress of the population – in contrast to the 1060s; this failure dramatically further undermined the authority of the Caliph.[199] The salvation for the Fatimids came through still loyal troops from the remaining bases on the coast of Palestine, which were under the command of the Armenian-born General Badr al-Ǧamālī. He marched to Cairo in January 1074 and was able to lure the rebelling commanders of the other regiments into a trap at a feast and had them all killed. In addition, he forced the merchants of grain and bakers to bring their hoarded stocks at low prices to the market of the capital by sheer physical threat, including the beheading of some of their representatives. Badr al-Ǧamālī (1074-1094) and his son al-Afdal (1094-1121) held the office of the vizir and the de facto power in Fatimid Egypt for almost the next 50 years.[200]

In any case, the "Great Calamity" would very much frustrate any coordinated Fatimid efforts to defend the provinces in Syria against the advance of the Seljuks from Mesopotamia. But also in Byzantine historiography, written from the "benefit" or "bias" of hindsight, the reigns of Constantine X Dukas and Romanos IV Diogenes seem inevitably to lead to the catastrophe of Manzikert in 1071 and the following loss of Asia Minor. This time, also Michael Attaleiates (writing in 1079/1080[201]) resorts to celestial phenomena (including Halley´s Comet in 1066, whose last sighting between August and September 989 Leo Deacon had mentioned as ill-omened, see above) and other portents as harbingers of imminent doom[202]:

> "Before this year, in the month of September of the second indiction, on the twenty-third of that month [23 September 1063], during the second watch of the night, there was a sudden powerful earthquake, more frightening than any that had happened before, and it began in the western regions. It was so great in magnitude that it overturned many houses, leaving only a few undamaged. (…) In the regions of Macedonia, the coastal cities suffered more on that night than the others, I mean Rhaidestos and Panion and Myriophyton [see the map in **fig. 3**], where whole sections of the walls collapsed to their very foundations along with many houses, and many people died. In the Hellespont, Kyzikos was especially struck, where the ancient Greek temple was also shaken and most of it collapsed. (…) From that time on and for two years earthquakes continued to occur sporadically at various times, leaving mortal men speechless in wonder. (…) After the two-year period, an earthquake occurred that was larger than the frequent aftershocks, but smaller than the initial one. It happened at Nicaea in Bithynia and brought almost total devastation and ruin to the place. Its most important and large churches - the one founded in honor of the Wisdom of the Word of God, which was also the cathedral, and the one of the Holy Fathers, where the Council of the most Holy and Orthodox Fathers against Areios confirmed its decisions and where Orthodoxy was proclaimed openly to shine brighter than the sun-those churches, then, were shaken and collapsed as did the walls of the city along with many private dwellings. And on that day the shaking ceased. These events were earned by our sins and were surely caused by divine anger; but it seems also that they were a predictive sign of the invasion by that nation, which I mentioned, and its destruction, for in divine signs it is possible to glimpse not only the things that we have already spoken about but also some things to come. During the course of the month of May of the fourth indiction [1066], a bright comet [Halley´s comet[203]] appeared after the sun had set, which was as large as the moon when it is full, and it gave the impression that it was spewing forth smoke and mist. On the following day it began to send out some tendrils and the longer they grew the smaller the comet became. These rays stretched toward the east, the direction toward which it was proceeding, and this lasted for forty days. From the month of October until the following May [1067], the emperor [Constantine X Dukas] was afflicted by illness, which wore him out and so he departed from this life."[204]

---

[198] al-Maqrīzī, *Ighāthah*, transl. Allouche, p. 37. Telelis, Μετεωρολογικά φαινόμενα, nr 480, 484, 487; Hassan, "Extreme Nile floods".

[199] Halm, *Die Kalifen von Kairo*, pp. 404–20; Ellenblum, *The Collapse of the Eastern Mediterranean*, pp. 151–55; Brett, *The Fatimid Empire*, pp. 201–06. In contrast, Chipman/Avni/Ellenblum, "Collapse, affluence, and collapse", 200–01, 204–05 and 210–11, propose a more or less climate-deterministic scenario, arguing that in the face of low Nile floods, "the relative capacity or lack of ability of the central government in these circumstances had very limited influence on the state's fortunes".

[200] al-Maqrīzī, *Ighāthah*, transl. Allouche, p. 39. Feldbauer, *Die islamische Welt*, pp. 361–62; Halm, *Kalifen und Assassinen*, pp. 17–86; Brett, *The Fatimid Empire*, pp. 210–15.

[201] Neville, *Guide to Byzantine Historical Writing*, pp. 150–52.

[202] Krallis, "Historiography as Political Debate".

Similar descriptions and prophetic notions can be found in Skylitzes continuatus, who follows Attaleiates in these passages.[205]

---

[203] On observations of Halley´s comet in 1066, which was even depicted on the famous Bayeux Tapestry, in other parts of Europe and the Mediterranean see Wozniak, *Naturereignisse im frühen Mittelalter*, pp. 107–12.
[204] Michael Attaleiates, *History* 15, 1–7, ed. and transl. Kaldellis/Krallis, pp. 160–67.
[205] John Scylitzes continuatus, *Chronicle*, ed. and transl. McGeer, pp. 64–67. See also Neville, *Guide to Byzantine Historical Writing*, p. 157; it is unclear if this continuation was written by John Skylitzes himself or by another (anonymous) author.

In contrast to Egypt, however, we hear few about climatic extremes; only during the first campaign of Romanos IV Diogenes, who succeeded on the throne in January 1068, against the Seljuks in the East, we hear from Michael Attaleiates as eyewitness that the Byzantine army suffered from cold conditions after it had crossed the Taurus mountains to the north in December 1068: "And proceeding through that country in this way and crossing over the Tauras mountain with his entire army, he [Romanos IV Diogenes] entered Roman territory. But the men marching with him, who were coming from a warm climate, suddenly found themselves in icy cold weather, with everything covered with frost. It was about the end of the month of December, and they felt the bitter cold. Thus it happened that horses, mules, and men, especially those whose bodies were not robust or well clothed, froze to death in the sudden cold and had to be left on the road, a pitiable sight."[206]

Occasion to reflect on the enmity of both Seljuks and nature against empire is provided for Michael Attaleiates, however, when Turkmen raiders plundered the city of Chonai (see the map in **fig. 3**) and its famous sanctuary of Archangel Michael in 1070:

> "Even before these news became widely known, they heard more tidings to the effect that the Turks had taken by storm the city of Chonai and the very shrine of the Archgeneral [Michael], famous for its miracles and dedications, and that they had filled the place with slaughter and filth, and polluted the church with many outrages. Worst of all was this: the channels in the cavern that, ever since the ancient visitation and divine manifestation of the Archgeneral, funnel the rivers flowing past that area whose current is precipitous, turbulent, and swift, failed to protect the refugees who sought to escape from the danger in them. Instead-and this had never happened before-the water flooded, was then sucked down, and again disgorged. It drowned almost all the fugitives, submerging them under water even though they were on land. The news of this greatly depressed us, for it was as though these disasters were being caused by divine anger. Not only the enemy but the very forces of nature seemed to be fighting against us."[207]

---

[206] Michael Attaleiates, *History* 17, 20, ed. and transl. Kaldellis/Krallis, pp. 218–19. Cf. also John Scylitzes continuatus, *Chronicle*, ed. and transl. McGeer, pp. 94–95. Telelis, Μετεωρολογικά φαινόμενα, nr 485. There are parallel reports on a severe winter for Central Europe, see Wozniak, *Naturereignisse im frühen Mittelalter*, pp. 500–01.

[207] Michael Attaleiates, *History* 19, 3–4, ed. and transl. Kaldellis/Krallis, pp. 256–257. On the event see also Thonemann, *The Maeander Valley*, p. 93. This passage is also adapted by Scylitzes continuatus; he augments it, however, with some interesting observations on the differences between earlier Seljuk attacks onto marginal provinces in the east with non–orthodox population and the recent raids into the Byzantine–orthodox heartlands: "For in times previous so great an invasion and onslaught of foreign enemies and the decimation of the people living under Roman rule were taken to be the wrath of God, but it was directed against those heretics who inhabit Iberia and Mesopotamia as far as Lykandos and Melitene, as well as the adjacent Armenia, or the ones who practise the Judaic heresy of Nestorios and the Akephaloi. These lands are full of this sort of erroneous belief. But when the calamity affected the Orthodox, all those who practised the faith of the Romans were at a loss as to what to do,

The severe crisis of the empire in the aftermath of the defeat at Manzikert of 1071 in Attaleiates´ narrative, however, largely unfolds without the "very forces of nature" – in a combination of civil wars, fights between independently acting warlords of Byzantine, Armenian, Norman or Turkmen background, devastations of large areas and displacements of masses searching refuge in the capital, for instance, a meltdown of state finances reflected in the debasement of the Byzantine gold coins, all aggravated by the incompetence or even viciousness of officials around the new young Emperor Michael VII Dukas (1071-1078), especially the eunuch Nikephoritzes (in some aspects a narrative revenant of the eunuch John Orphanotrophos).[208]

In addition to his criticism of the regime of Michael VIII, Michael Attaleiates, however, did not abstain from other unhappy omens in his narrative:

> "In that year a number of portents were observed in the City of Byzas [Constantinople]. A three-legged chicken was born as well as a baby with an eye on its forehead (and having a single eye at that) and the feet of a goat. When it was exposed in the public avenue in the area of Diakonissa, it uttered the cries of a human baby.[209] Two soldiers of the Immortals [a guard regiment] were struck by lightning in a public place close to the western walls of the City. Not only that but certain comets streaked across the sky. Meanwhile, because the east was being wasted by the barbarians there who were ruining and subjecting it, large multitudes were fleeing those regions on a daily basis and seeking refuge in the Imperial City [Constantinople], so that hunger afflicted everyone, oppressing them because of the lack of supplies. When winter arrived, because the

---

thinking that they had reached their limit of iniquity just as the Amorites had reached theirs, and believing that in these circumstances not only correct belief was required but also a life not at odds with faith. It followed that both the man who clearly erred with respect to belief and the man who stumbled and fell into an imperfect way of life were subject to the same punishment. Whosoever practises this and teaches this is praised and blessed." John Scylitzes continuatus, *Chronicle*, ed. and transl. McGeer, pp. 108–09.

[208] Krallis, *Michael Attaleiates and the Politics of Imperial Decline*, pp. 171–212; Krallis, "Historiography as Political Debate"; Kaldellis, *Streams of Gold*, pp. 248–54; Preiser-Kapeller, "Byzantium 1025–1204", pp. 63–64; Hendy, *Studies in the Byzantine Monetary Economy*, pp. 233–36, 509–12; Caplanis, "The Debasement of the "Dollar of the Middle Ages""; For the background to the "loss" of Anatolia see now especially Beihammer, *Byzantium and the Emergence of Muslim-Turkish Anatolia*, pp. 9, 15, 388 and 390, who sums up: "The political situation of Byzantine Asia Minor from 1056 onwards was marked by serious tensions between centralizing tendencies and the gradual strengthening of regional powers backed by military forces. These consisted of seditious Byzantine aristocrats, foreign mercenary troops, Armenian noblemen, Arab and Kurdish emirs and many others. This process resulted in a fragmentation of state authority and the emergence of numerous, mostly short–lived, semi–independent local lordships of limited size. Political power, to a large extent, was regionalized. This is to say, that we are dealing not necessarily with a conflict between the Byzantine central government and Turkish invaders but with struggles and contentions within a complicated patchwork of local powers, in which the Turks intruded and eventually managed to prevail."

[209] One is reminded of a similar portent reported by Bar Hebraeus for Baghdad in 1031, see Bar Hebraeus, *Chronography*, transl. Budge, pp. 193–94, and above. On portents in Attaleiates see also Krallis, *Michael Attaleiates and the Politics of Imperial Decline*, pp. 205–10.

emperor [Michael VII Dukas] lacked generosity and was extremely stingy, he offered no succour from the imperial treasuries or any other form of provident welfare either to those in office or to the people of the City, and so each person wallowed in his own misery, nor did he hold out an abundant hand that could assist the poor and provide them with daily provisions, for it is through these means that the poor are normally supplied with necessities. There were many, indeed countless deaths every day, not only among the refugees but also among the people of the City [Constantinople], so that their dead bodies were heaped both in the so-called porticos and in open spaces, and they were carried on stretchers, each one of which was often stacked with five or six bodies piled up in a random heap. Everywhere you saw sad faces and the Reigning City [Constantinople] was filled with misery. The rulers did not let up on their daily injustices and illegal trials, but acted as though the Romans were not being afflicted by anything out of the ordinary, be it foreign war, divine wrath, or poverty and violence oppressing the people; it was with such nonchalance that they practiced all their tyrannical impieties. Every imperial scheme and plan, in fact, was preoccupied with some injustice against their own subjects, at the ingenious looting of their livelihoods and their resources for living."[210]

The reference to the porticoes reminds at least the modern reader of the winter of 927, when Romanos I Lakapenos used them as emergency shelter for the homeless of the capital, while now they served for the deposition of the bodies of the victims of the famine, according to Attaleiates aggravated by the "nonchalance" and "tyrannical impieties" of the current government. Another example of Attaleiates for the aggravation or even causation of supply shortfalls by "imperial plans" is the establishment of the so-called *phoundax* by the *logothetes tou dromou* Nikephoritzes in the important grain market of Rhaidestos (see the map in **fig. 3**) in Thrace. This according to Michael Attaleiates contributed to a scarcity of food and an increase of prices in Constantinople, in turn raising resistance against the regime of the emperor.

"He [Nikephoritzes] thereby established a monopoly over this most essential of trade, that of grain, as no one was able to purchase it except from the *phoundax*. (…) For from that moment on they monopolized not only the grain carts (…) but also all other goods the circulated in the vicinity. (…) He, then, farmed out the phoundax for sixty pounds of gold, and enjoyed the proceeds, while everyone else was hard-pressed by a shortage not only of grain but of every other good. For the dearth of grain causes dearth in everything else, as it is grain that allows the purchase or preparation of other goods, while those who work for wages demand higher pay to compensate for the scarcity of food. (…) As a result of the emperor´s planning or, rather, of Nikephoros´s evil designs, grain was in short supply and abundance turned into dearth. The people´s discontent increased."[211]

---

[210] Michael Attaleiates, *History* 26, 8–9, ed. and transl. Kaldellis/Krallis, pp. 384–87. See also John Scylitzes continuatus, *Chronicle*, ed. and transl. McGeer, pp. 162–65; Bar Hebraeus, *Chronography,* transl. Budge, p. 226. Wozniak, *Naturereignisse im frühen Mittelalter*, pp. 625–26.

While Michael Attaleiates provides interesting observations on the impact of a price increase in basic foodstuffs on other economic sectors, we have to be aware that he is not necessarily a trustworthy rapporteur. During the reign of Michael VII Dukas, he had no qualms to cooperate with the regime and to seek favours from it; his highly critical narrative was written only after the downfall of Michael VII and Nikephoritzes, also to create an evil counterexample to the new emperor Nikephoros III Botaneiates (1078-1081), to whom he dedicated his text. Furthermore, Michael Attaleiates himself was engaged in the grain trade at Rhaidestos as landowner in the city´s environs and proprietor of a landing place for cargo vessels in the capital and therefore affected by the introduction of the *phoundax*, whose impact on the prices also arose from the decision of market actors such as Attaleiates to withhold their stocks of grain under these less profitable circumstances.[212]

Nevertheless, it seems plausible that a majority of the population in Constantinople held the regime of Michael VII Dukas responsible for the sorry state of the empire and especially of the supplies of the capital, which lead to the eventual downfall of the government and the acceptance of Nikephoros III as emperor.[213] In contrast to Ellenblum´s scenario, however, no "short-term climatic catastrophe" can be identified from the written sources as contributing either to the regime chance of 1078 or to the severe crisis of the Byzantine Empire in the 1070s or 1080s; on the occasion of another rebellion in the European provinces during the reign of Michael VII, we learn from Attaleiates that when the "summer was beginning (…) the fruit was still hanging unpicked from the trees", since people had fled from their farms due to ongoing warfare; therefore, "there was not small shortage of food in the Reigning City and the other western cities".[214] So there was no lack of rain, temperature or vegetation, but of security.

---

[211] Michael Attaleiates, *History* 25, 4–6, ed. and transl. Kaldellis/Krallis, pp. 366–73: For interpretations of this passage: Angold, *The Byzantine Empire*, pp. 122–23; Laiou/Morrisson, *The Byzantine Economy*, pp. 135–36; Laiou, "God and Mammon"; Dölger/Wirth, *Regesten von 1025–1204*, nr 996. See also John Scylitzes continuatus, *Chronicle*, ed. and transl. McGeer, pp. 146–47.
[212] Krallis, *Michael Attaleiates and the Politics of Imperial Decline*, pp. 20–29; Kaldellis, *Streams of Gold*, pp. 262–63; Krallis, "Historiography as Political Debate".
[213] Kaldellis, *Streams of Gold*, pp. 265–66; Preiser-Kapeller, "Byzantium 1025–1204", pp. 63–64.
[214] Michael Attaleiates, *History* 26, 5, ed. and transl. Kaldellis/Krallis, pp. 378–81.

From neighbouring regions, we read that in 1074 in Syria the effects of ongoing warfare on the food situation around Damascus were aggravated by a plague of rodents, resulting in famine and disease.[215] More generally, however, Matthew of Edessa connects the "severe famine" which in 1079/1080 "occurred throughout all the lands of the venerators of the cross, lands which are located on this side of the Mediterranean Sea" to the raids of the "bloodthirsty and ferocious Turkish nation", since they "interrupted the cultivation of land", causing a "shortage of food"; furthermore, "the cultivators and laborers decreased due to the sword and enslavement, and so famine spread throughout the whole land. Many areas became depopulated, the Oriental peoples began to decline, and the country of the Romans became desolate; neither food nor security for the individual was to be found anywhere except in Edessa and its confines."[216] Again, we do not read about direct climatic factors. Baghdad, however, between November 1073 and February 1074 suffered from torrential winter rains and severe floods, destroying large areas in the east of the city.[217]

**11 The Komnenian dynasty, the end of the Oort Minimum and the Medieval Climate Anomaly in the Eastern Mediterranean in the 12th to 13th century**

In 1081, the elderly Nikephoros III Botaneiates was forced to step down from the throne in favour of the young Alexios I Komnenos[218]; the later pro-Komnenian historiography, especially the "Alexias" written down by Alexios´ daughter Anna in the 1140s, celebrated him as saviour of the state, stabilizing the empire in terms of administration, finances and especially its borders, first on the Balkans, then in Asia Minor.[219] We have a contrary opinion in the 12th century history of Zonaras, who moaned that Alexios I "did not behave with (the state´s) institutions as common and as with public ones and did not consider himself their steward but their owner; he considered and named the palace his own house. (…) he offered the relatives and some servants public funds in whole wagonloads, and gave them ample annual allowances, so that they embraced themselves with great wealth."[220]

While earlier scholarship has followed this opinion and has characterised Alexios I reign as completion of the takeover of the state by the powerful families struggling for access to the throne since the 10th century or even as a decisive step toward a "feudalisation" of Byzantium, recent research has demonstrated, that Alexios I did not dissolve the "centralised" Roman polity of the past, but even intensified already existing instruments for the grasp of the state onto taxes and properties created by the Macedonian legislation (see above). Thereby, the Byzantine government re-established the material basis to act as great power also in the 12th century (especially in the reign of Alexios´ I grandson Manuel I, 1143-1180) with a radius of action still comparable with the one of the assumed singular apex of Byzantine power in the early 11th century (under Basil II), ranging from Southern Italy to the Euphrates and from beyond the Danube (Hungary) to Egypt (target of naval operations in the 1160s and 1170s).[221]

---

[215] Bar Hebraeus, *Chronography*, transl. Budge, pp. 225–26.
[216] Matthew of Edessa, *History* II, 73, transl. Dostourian, p. 143.
[217] Bar Hebraeus, *Chronography*, transl. Budge, pp. 224–25. Telelis, *Μετεωρολογικά φαινόμενα*, nr 491.
[218] Kaldellis, *Streams of Gold*, pp. 269–70; Preiser-Kapeller, "Byzantium 1025–1204", pp. 64–65.
[219] Neville, *Guide to Byzantine Historical Writing*, pp. 174–79; Kaldellis, *Streams of Gold*, pp. 287–301; Preiser-Kapeller, "Byzantium 1025–1204", pp. 67–70.
[220] Zonaras, *Epitome* 18, 29, ed. Büttner-Wobst, p. 767. It is unclear if John Zonaras wrote his history also in response to Anna´s *Alexias*, see Neville, *Guide to Byzantine Historical Writing*, p. 195.

Regarding climatic events, besides localised events such as stormy weathers on the sea between Greece and Southern Italy during warfare with the Normans in 1081 and 1085[222], Anna Komnene for Alexios´ reign refers to extreme conditions in the winter 1090/1091, when after a defeat against the Pechenegs and due to a threat to the maritime supply lines of Constantinople by the fleets of the Turkish Emir Tzachas of Smyrna the situation for Alexios I was already precarious: "Things were not going well for Alexios either by sea or on land, and the severe winter did not help; in fact, the doors of houses could not be opened for the heavy weight of snow (more snow fell that year than anyone could remember in the past). Still, the emperor did all that he could by summoning mercenaries by letter from all quarters."[223]

This extreme winter overlaps with reports on climate anomalies such as severe winters and dry summers in Western Europe in the early 1090s, followed by famines and epidemics among humans and animals, which have been discussed as possible motivations for the mobility of laymen toward a "promised land", where "milk and honey flows" in the First Crusade after 1095.[224] For Northern Iraq, the Kuna Ba data indicates the years between 1085 and 1094 as the most humid ones in the entire 11th century (see **fig. 6**).[225] In addition, very cold winters with heavy snowfall were reported for 1091 and 1093 in China.[226]

For John Zonaras, on the contrast, like earlier governments for Skylitzes or Attaleiates, also the reign of Alexios I Komnenos in general was beclouded by portents and catastrophes:

---

[221] Smyrlis, "Private property and state finances"; Smyrlis, "The Fiscal Revolution of Alexios I Komnenos"; Magdalino, *The Empire of Manuel I Komnenos*; Kaldellis, *Streams of Gold*, pp. 276–277; Preiser-Kapeller, "Byzantium 1025–1204", pp. 73–76; Olson, *Environment and Society in Byzantium*, pp. 193–94.
[222] Telelis, Μετεωρολογικά φαινόμενα, nr 494, 496–497.
[223] Anna Komnene, *Alexias* VIII, 3, ed. Reinsch/Kambylis, pp. 241, 78–242, 84; transl. Sewter/Frankopan, p. 220. Telelis, Μετεωρολογικά φαινόμενα, nr 502. Cold conditions are also indicated for Albania in the early 1090s in tree ring data, see PAGES 2k Network consortium, Database.
[224] i Monclús, "Famines sans frontiers", pp. 48–50; Wozniak, *Naturereignisse im frühen Mittelalter*, pp. 514–15, 533–34, 545, 627–28, 658–60, 685; Salvin, "Crusaders in Crisis"; Pfister/Wanner, *Klima und Gesellschaft in Europa*, pp. 172–74.
[225] Sinha et al., "Role of climate".
[226] Zhang, *The River, the Plain, and the State*, p. 156.

> "During the reign of this emperor [Alexios I Komnenos], there were many fires in different parts of the city [of Constantinople], and many of them the fire ravaged and destroyed. Once a very strong and violent wind blew under this ruler during the springtime, which caused many structures to collapse. The statue, which was placed on the large round porphyry column on the Plakaton, fell down and killed many passers-by. The image, which broke and fell into many pieces when it fell, was of immense size and wonderful beauty. Another time a very heavy downpour broke out, just on the festive day of the chief apostles of Christ, Peter and Paul; it began late in the evening and lasted until the same hour of the following day without subsiding. At that time houses collapsed due to the rush of water, the valleys were filled with water and were in no way different from seas, and not a few people and many animals died."[227]

Again, the number of calamities and climatic events we find in a source also depends on its narrative intentions such as its possible framing of an emperor´s reign through the lens of "moral meteorology". Ellenblum was right that we cannot subordinate historiography to the primacy of the proxy data, but we also cannot take its information on the frequency, severity, or impact of environmental stressors at face value. Nevertheless, written evidence from various points of observations (Constantinople, Baghdad, Cairo, Armenia) can allow us to "triangulate" actual clusters of climatic extremes across the Eastern Mediterranean and the Middle East (or with a glance on the rich "archives of society" in Western Europe or China even across Afro-Eurasia), as during the Oort Solar Minimum. In some cases, high-resolution proxies provide additional confirmation.

Regarding the interplay between environmental and socio-politic dynamics, however, the texts we have inspected themselves often provide more complex lines of causation beyond mere linear climatic impacts on a society, hinting at the role of institutions and market actors for the mitigation or aggravation of a supply shortfall after a strong winter or a low Nile flood, for instance. And beyond the focus on "short-term climatic catastrophes" propagated by Ellenblum and his co-authors also in his last publications[228], we have to consider the interplay between long term trends in climate, agriculture and society and singular shocks in order to provide convincing scenarios which do justice both to the possible fragility as well as resilience of regimes and polities during the Medieval Climate Anomaly.

---

[227] Zonaras, *Epitome* 18, 26, ed. Büttner-Wobst, pp. 755–56. Telelis, Μετεωρολογικά φαινόμενα, nr 495.
[228] Li/Shelach–Lavi/Ellenblum, "Short–Term Climatic Catastrophes"; Chipman/Avni/Ellenblum, "Collapse, affluence, and collapse".

After the end of the Oort Solar Minimum around 1080/1090, more stable temperature and precipitation conditions returned in Western and Central Europe in the 12th and first half of the 13th centuries, of course not excluding short term anomalies and calamities such as a "mega drought" in 1137.[229] "Beneficial" climate parameters can be reconstructed equally for parts of the Eastern Mediterranean in the 12th century; more humid conditions as well as an increase of human activity during the Crusader period has been identified in proxy data from the site of Jableh, then within the territory of the Principality of Antioch (see the map in **fig. 3**). A similar increase in indicators of human agricultural activity can be seen in pollen data from the Southern Bekaa Valley in Lebanon. Besides these long-term trends, however, again written evidence documents the recurrence of severe droughts, famine and plague of locusts and rodents in the Crusader states especially also during the 12th century; but in general, the "Latin" regimes were able to cope with these situations better than in the 13th century, when also general political-military conditions had changed to their disadvantage.[230]

Equally, for the remaining Byzantine provinces at the Aegean, especially Greece, various types of evidence confirms the scenario of continued or even intensified 12th century economic growth, be it the increase in the number of settlements mentioned in the documents of Mount Athos for the Chalkidike, the increase in the number of church buildings in Messenia from the 10th to the 13th century or the increase in the monetary finds in Corinth or Athens.[231] Even the increase in the number of sites of Venetian commercial activity (as documented in the Chrysobulls of 1082 and 1198) especially in the provinces of the Southern Balkans can be interpreted as indicator of economic growth and demand – despite the long term consequences these activities may have had.[232]

---

[229] Goosse et al., "The medieval climate anomaly in Europe"; Sirocko/David, "Das mittelalterliche Wärmeoptimum"; Behringer, *Kulturgeschichte des Klimas*, pp. 103–15; i Monclús, "Famines sans frontiers"; Pfister/Wanner, *Klima und Gesellschaft in Europa*, pp. 175–84.

[230] Kaniewski et al., "Medieval coastal Syrian vegetation patterns"; Kaniewski et al., "The Medieval Climate Anomaly and the Little Ice Age in Coastal Syria"; Hajar et al., "Environmental changes in Lebanon" (for the Southern Bekaa–Valley). Cf. also Redford, "Trade and Economy in Antioch and Cilicia". On famine and droughts in the Crusader states and their reactions to it cf. Raphael, *Climate and Political Climate*, pp. 21–27, 56–94 and 191–93; Xoplaki et al., "Modelling Climate and Societal Resilience".

[231] Kazhdan/Epstein, *Change in Byzantine Culture*, pp. 34–37; Angold, *The Byzantine Empire*, pp. 280–86; Laiou/Morrisson, *The Byzantine Economy*, pp. 91–96; Whittow, "The Middle Byzantine Economy", pp. 475–76; Curta, *Southeastern Europe in the Middle Ages*, pp. 323–27; Gerolymatou, Αγορές, έμποροι και εμπόριο, pp. 152–70; Bintliff, *The Complete Archaeology of Greece*, pp. 391–93 (with graphs); Izdebski/Koloch/Słoczyński, "Exploring Byzantine and Ottoman economic history"; Olson, *Environment and Society in Byzantium*, p. 196.

[232] Lilie, *Handel und Politik*, esp. pp. 117–221 on the Italian presence in the cities in the Byzantine provinces; Laiou/Morrisson, *The Byzantine Economy*, pp. 141–47; Gerolymatou, Αγορές, έμποροι και εμπόριο, pp. 102–09; Whittow, "The Middle Byzantine Economy", pp. 476–77; Jacoby, "Venetian commercial expansion".

These long-term trends even continued when we can reconstruct another change of climatic conditions towards less "favourable" parameters in the Eastern Mediterranean and Middle East from the middle to the end of the 12th century onwards. In Cappadocia (now beyond the Byzantine borders), the oxygen isotope data from Lake Nar indicates a trend reversal towards more arid conditions already from the early 11th century onwards, with the driest period in the 1180s (see **fig. 8**).[233] For North-western Anatolia, proxy data (speleothems from the Sofular cave, see **fig. 7**) documents a turn towards more arid conditions from the 1180s onwards, which continued until after the mid-13th century.[234] A similar pattern can be identified in a May-June precipitation reconstruction based on tree rings for the Northern Aegean (see **fig. 9**).[235] In another precipitation reconstruction for South-western Anatolia for the years 1097 to 2000, the 70 years from 1195 to 1264 were identified as the driest period in the entire record (while the years 1098 to 1167 marked one of the most humid ones). An equal pattern can be found in speleothems from Thrace (Uzuntarla Cave, Turkey), in sediments of the Tecer Lake from Cappadocia (see the map in **fig. 3**) or in the pollen data from the area of Antioch, with a shift towards drier conditions from the later 12th century onwards, which overlapped with a severe famine in Syria between 1178 and 1181.[236] A similarly dry period is indicated in the Kuna Ba data from Northern Iraq, starting around 1165 and continuing until the mid-13th century (see **fig. 6**).[237] Around 1180, a 30 years periods of severe El Niño events began, and Egypt, now under Ayyubid rule, was hit by extremely low Nile floods in the years around 1200; the catastrophic famine which followed and the horrendous social phenomena accompanying it have been described by an eyewitness, the physician and historian ʿAbd al-Laṭīf from Baghdad, who spent these years in Egypt.[238] In various regions, these climatic change to more unstable, arid and/or cold conditions continued beyond the mid-13th century already into the general period of transition from the Medieval Climate Anomaly toward the "Little Ice Age" across Afro-Eurasia, which we have discussed for Byzantium in an earlier publication.[239]

## 12 Conclusion and outlook

---

[233] Woodbridge/Roberts, "Late Holocene climate"; Dean et al., "Palaeo-seasonality of the last two millennia".
[234] Fleitmann et al., "Sofular Cave".
[235] Griggs et al., "A regional high–frequency reconstruction".
[236] Kuzucuoglu et al., "Mid- to late-Holocene climate change". A more detailed discussion and references to this evidence can be found in Preiser-Kapeller, "A Collapse of the Eastern Mediterranean?".
[237] Sinha et al., "Role of climate".
[238] Grove/Adamson, *El Niño in World History*, pp. 52–55; Nash et al., "African hydroclimatic variability". ʿAbd al-Laṭīf al-Baġdādī, *Description of Egypt*, ed. and transl. Mackintosh-Smith, pp. 122–79. See also al-Maqrīzī, *Ighāthah*, transl. Allouche, pp. 41–42. Telelis, Μετεωρολογικά φαινόμενα, nr 595–597; Hassan, "Extreme Nile floods".
[239] Preiser-Kapeller/Mitsiou, "The Little Ice Age and Byzantium".

In sum, thus, from a palaeoclimatological perspective and synthesising important parts of the archives of society and of nature, we encounter a rather short and temporarily as well as spatially incoherent "Medieval Climate Anomaly" in the Eastern Mediterranean, punctuated by recurring anomalies and extreme events, but nevertheless not necessarily causing "collapse" in Byzantium or neighbouring polities, even correlating with a long-term "economic expansion", for certain times, at least. What we lack to complete and further nuance this picture are in-depth regional studies, ideally again combining rich written evidence (especially charters) with archaeological and proxy data (for both agricultural activities as well as temperature and precipitation conditions).

Recently, promising studies along these lines for the hinterlands of Ephesus (with pollen data from Lake Belevi, see **fig. 3**)[240] and of Miletus (see **fig. 3**, including pollen data from Bafa Gölü)[241] have been published. In addition to long term trends, such regional archives can informs us also on short-term localised events as in a charter for Andronikos Dukas issued by his cousin Emperor Michael VII in 1073, by which he granted Andronikos a large domain in the Maeander delta plain near Miletus; we learn that one of the estates, Mandraklou, had been damaged by a flood of the river before, so that of the 185 modioi only 36 were cultivable now, while the rest was marshland.[242] Ultimately, a combination of these spatial levels of analysis may allow us to describe in more detail the resilience (or fragility) of the Byzantine Empire in the face of climatic, epidemic and other exogenous shocks through the dynamics of its rural and urban communities.[243]

---

[240] Stock et al., "Human–environment interaction in the hinterland of Ephesos".
[241] Niewöhner et al., "The Byzantine Settlement History of Miletus"; Olson, *Environment and Society in Byzantium*, pp. 179–80.
[242] Βυζαντινά έγγραφα της Μονής Πάτμου: Β΄, ed. Nystazopoulou-Pelekidou, nr 50, lns. 270–274. Dölger/Wirth, *Regesten von 1025–1204*, nr 994; Thonemann, *The Maeander Valley*, pp. 259–70, esp. 266, and 302–06, on the further development of Mandraklou.
[243] See for instance also the "Palaeo-Science and History"-project of Adam Izdebski at the Max Planck-Institute for the Science of Human History in Jena, https://www.shh.mpg.de/1056512/byzres.

**Bibliography**

**Primary sources**

Novels of the Macedonian emperors, transl. E. McGeer, *The Land Legislation of the Macedonian Emperors* (Medieval Sources in Translation 38), Toronto 2000.

Symeon Magistros, ed. St. Wahlgren, *Symeonis Magistri et Logothetae Chronicon* (Corpus Fontium Historiae Byzantinae 44/1), Berlin 2006.

Theophanes Continuatus VI, ed. I. Bekker, *Theophanes continuatus, Joannes Cameniata, Symeon Magister, Georgius Monachus*, Bonn 1838.

Yaḥyā of Antioch, *Chronicle*, transl. B. Pirone, *Yaḥyā al-Anṭākī, Cronache dell'Egitto fāṭimide e dell'impero bizantino (937-1033)*, Bologna 2018.

Zonaras, *Epitome*, ed. Th. Büttner-Wobst, *Ioannis Zonarae Epitomae Historiarum Libri XIII–XVIII*, Bonn 1897

**Secondary Literature**

Adamson, G. C. D./Nash, D., "Climate History of Asia (Excluding China)", in S. White/Ch. Pfister/F. Mauelshagen (eds.), *The Palgrave Handbook of Climate History*, London 2018, pp. 203–11.

Andrews, T.L., *Mattʿēos Uṙhayecʿi and His Chronicle. History as Apocalypse in a Crossroads of Cultures*, Leiden 2017.

Andriollo, L., *Constantinople et les provinces d'Asie Mineure, IXe–XIe siècle. Administration impériale, sociétés locales et rôle de l'aristocratie*, Paris 2017.

Andriollo, L./Métivier, S., "Quel rôle pour les province dans la domination aristocratique au XIe siècle?", in B. Flusin/J.-Cl. Cheynet (eds.), *Autour du Premier humanisme byzantin & des Cinq études sur le XIe siècle, quarante ans après Paul Lemerle* (Travaux et Mémoires 21/2), Paris 2017, pp. 505–30.

Angold, M., *The Byzantine Empire 1025–1204*, London 1997.

Barthélemy, D., "L´aristocracie franque du XIe siècle en constraste avec l´aristocratie byzantine", in B. Flusin/J.-Cl. Cheynet (eds.), *Autour du Premier humanisme byzantin & des Cinq études sur le XIe siècle, quarante ans après Paul Lemerle* (Travaux et Mémoires 21/2), Paris 2017, pp. 491–504.

Bauer, Th., *Warum es kein islamisches Mittelalter gab. Das Erbe der Antike und der Orient*, Munich 2018.

Behringer, W., *Kulturgeschichte des Klimas. Von der Eiszeit bis zur globalen Erwärmung*, Munich 2007.

Beihammer, A. D., *Byzantium and the Emergence of Muslim-Turkish Anatolia, ca. 1040–1130*, London/New York 2017.

Bianquis, Th., "Une crise frumentaire dans l'Égypte Fatimide", *Journal of the Economic and Social History of the Orient* 23 (1980), 67–101.

Bintliff, J., *The Complete Archaeology of Greece: From Hunter-Gatherers to the 20th Century AD*, Malden/Oxford 2012.

Brandes, W., "Byzantine Predictions of the End of the World in 500, 1000, and 1492 AD", in H.-C. Lehner (ed.), *The End(s) of Time(s). Apocalypticism, Messianism, and Utopianism through the Ages*, Leiden 2021, pp. 32–63.

Brett, M., *The Fatimid Empire*, Edinburgh 2017.

# Figures

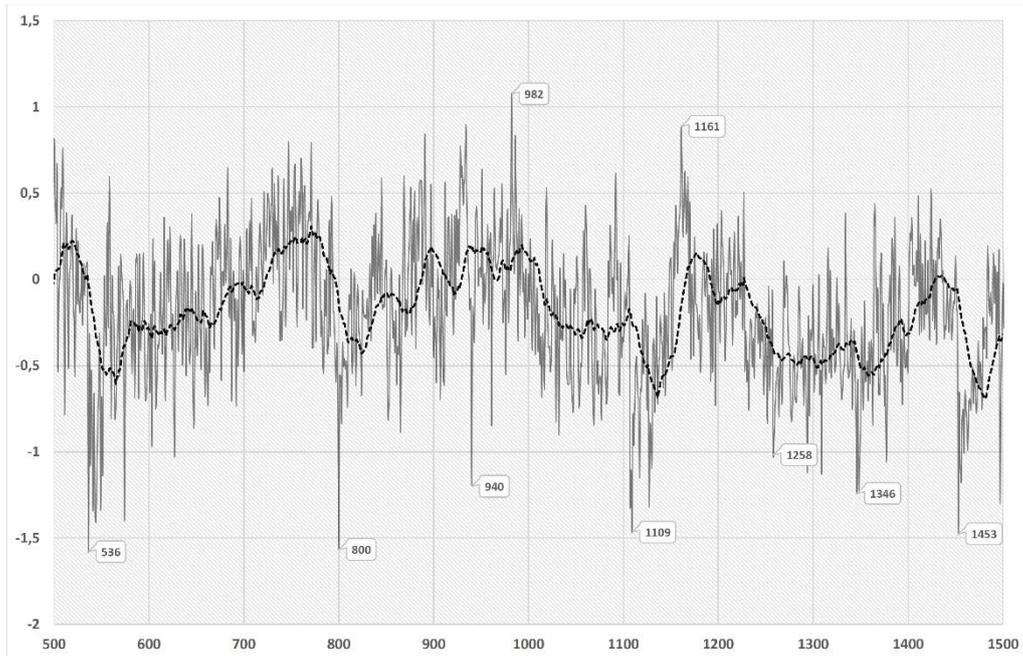

Fig. 1 Average summer temperatures in Western and Central Europe 500–1500 CE, reconstructed on the basis of tree rings (data: Luterbacher et al., "European summer temperatures"; graph: J. Preiser-Kapeller, OEAW, 2022)

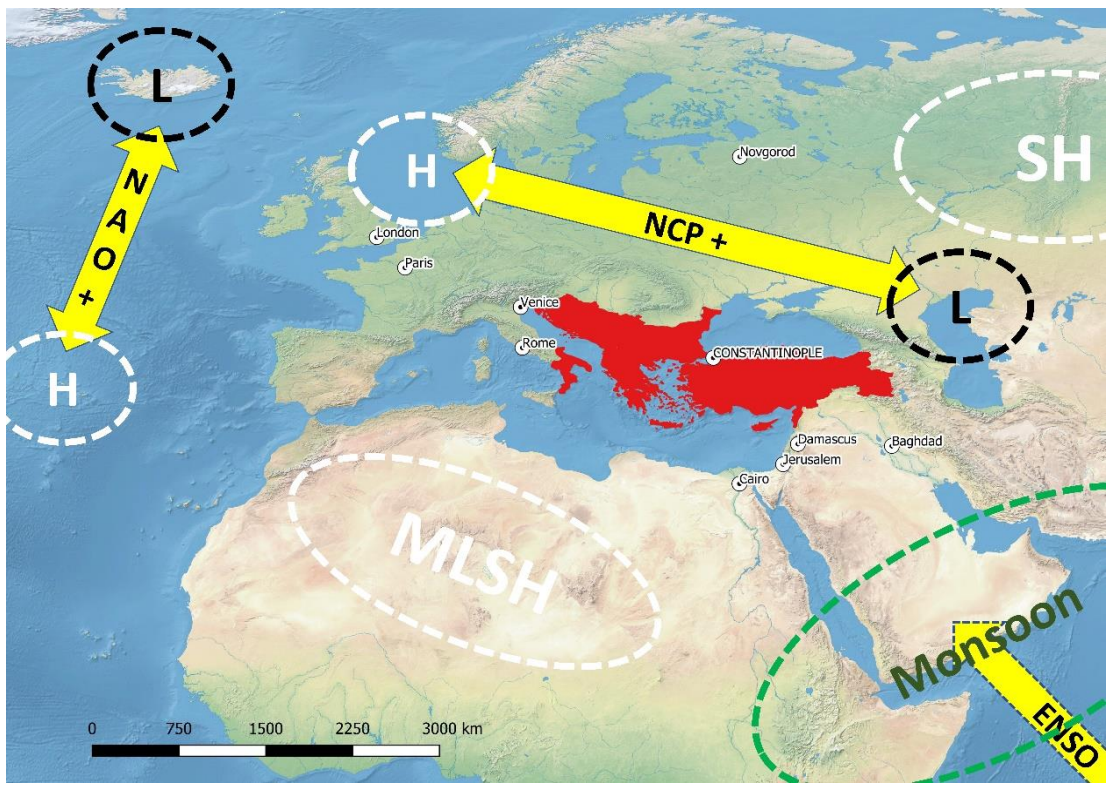

Fig. 2 Map of selected climate patterns influencing weather dynamics in the Eastern Mediterranean: NAO (North Atlantic Oscillation), NCP (North Sea-Caspian-Pattern), SH (Siberian High), ENSO (El Niño-Southern Oscillation), MLSH (Mid-Latitude Subtropical High-Pressure Systems); in red colour territorial extent of the Byzantine Empire in ca. 1045 CE (J. Preiser-Kapeller, OEAW, 2022)

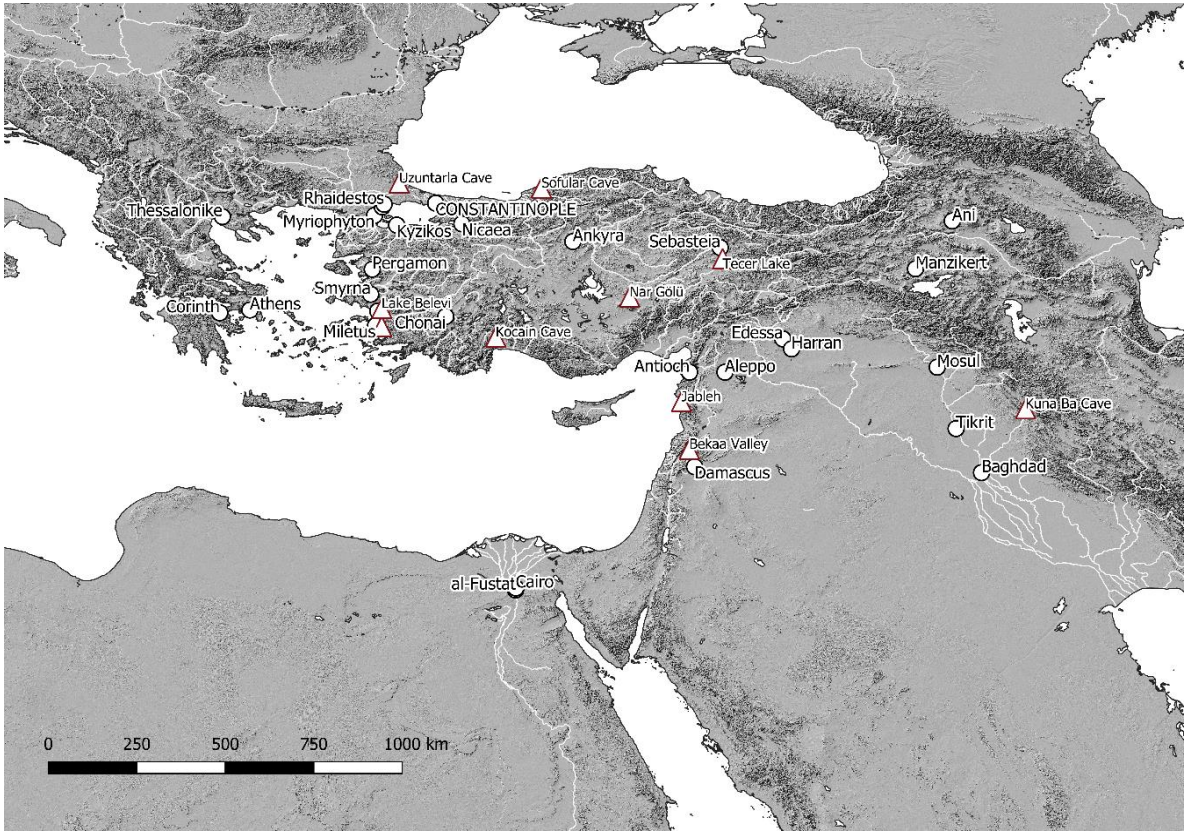

Fig. 3 Map of selected cities and towns (circles) and proxy data sites (triangles) mentioned in the chapter (J. Preiser-Kapeller, OEAW, 2022)

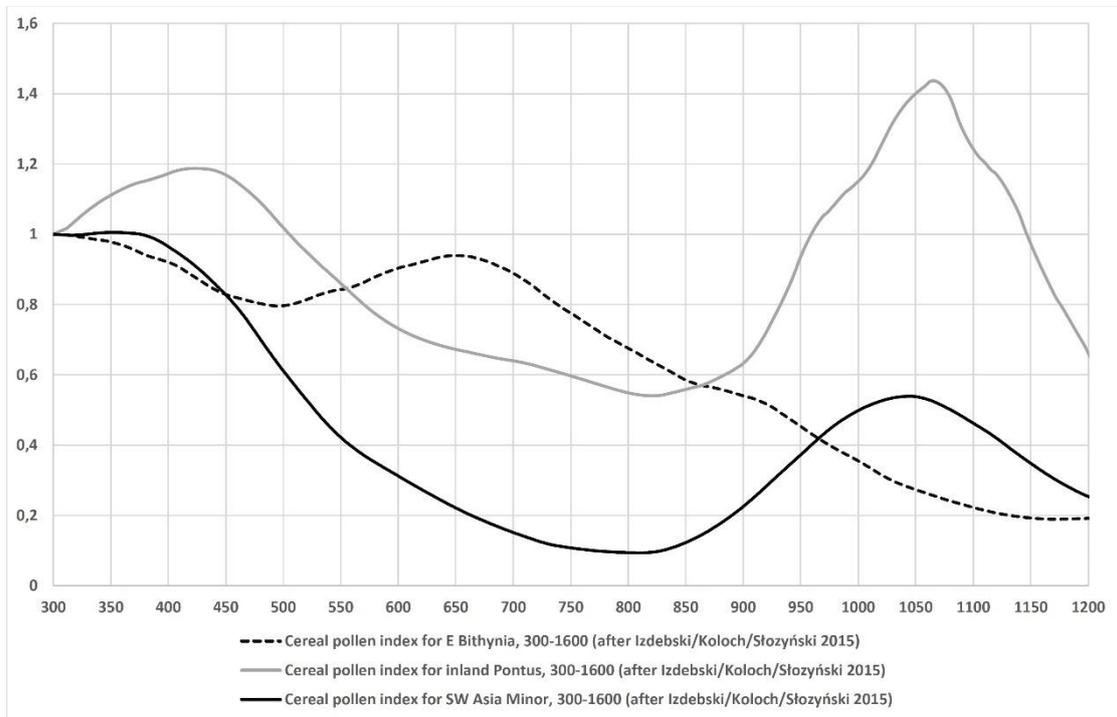

Fig. 4 Cerealia pollen data indices for three regions in Asia Minor, 300–1200 CE (data: Izdebski/Koloch/Słoczyński, "Exploring Byzantine and Ottoman economic history"; graph: J. Preiser-Kapeller, OEAW, 2022)

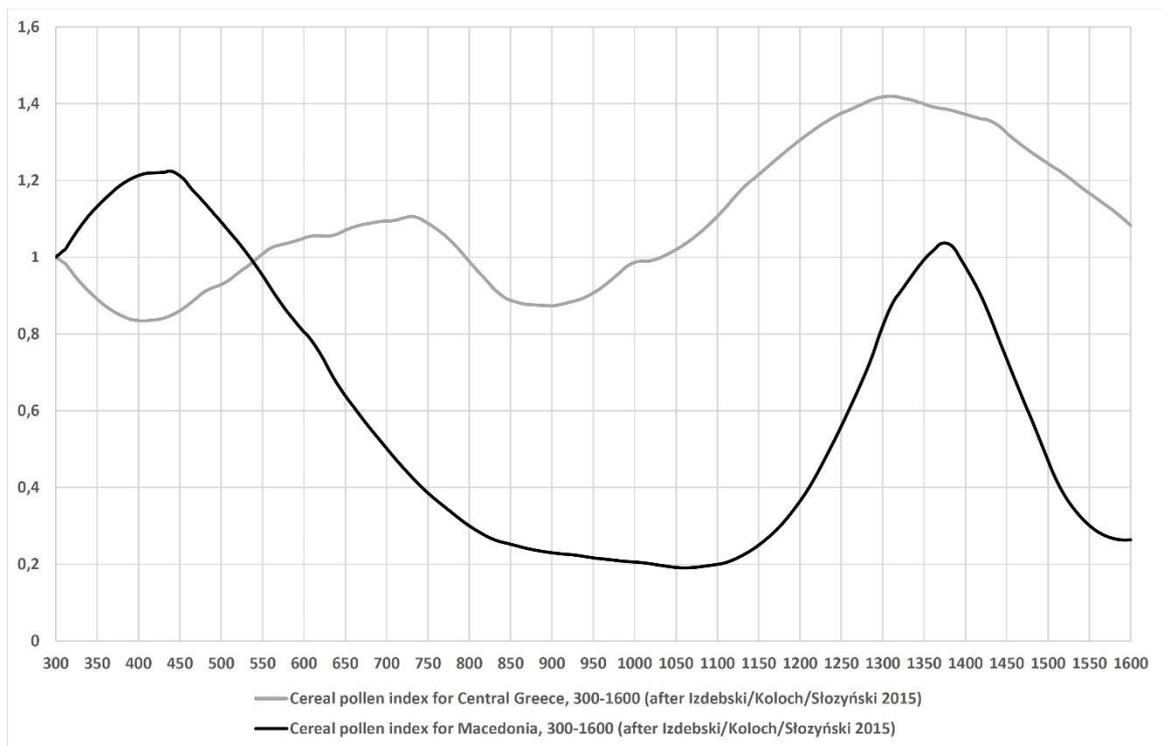

Fig. 5 Cerealia pollen data indices for two regions in Greece, 300–1600 CE (data: Izdebski/Koloch/Słozyński, "Exploring Byzantine and Ottoman economic history"; graph: J. Preiser-Kapeller, OEAW, 2022)

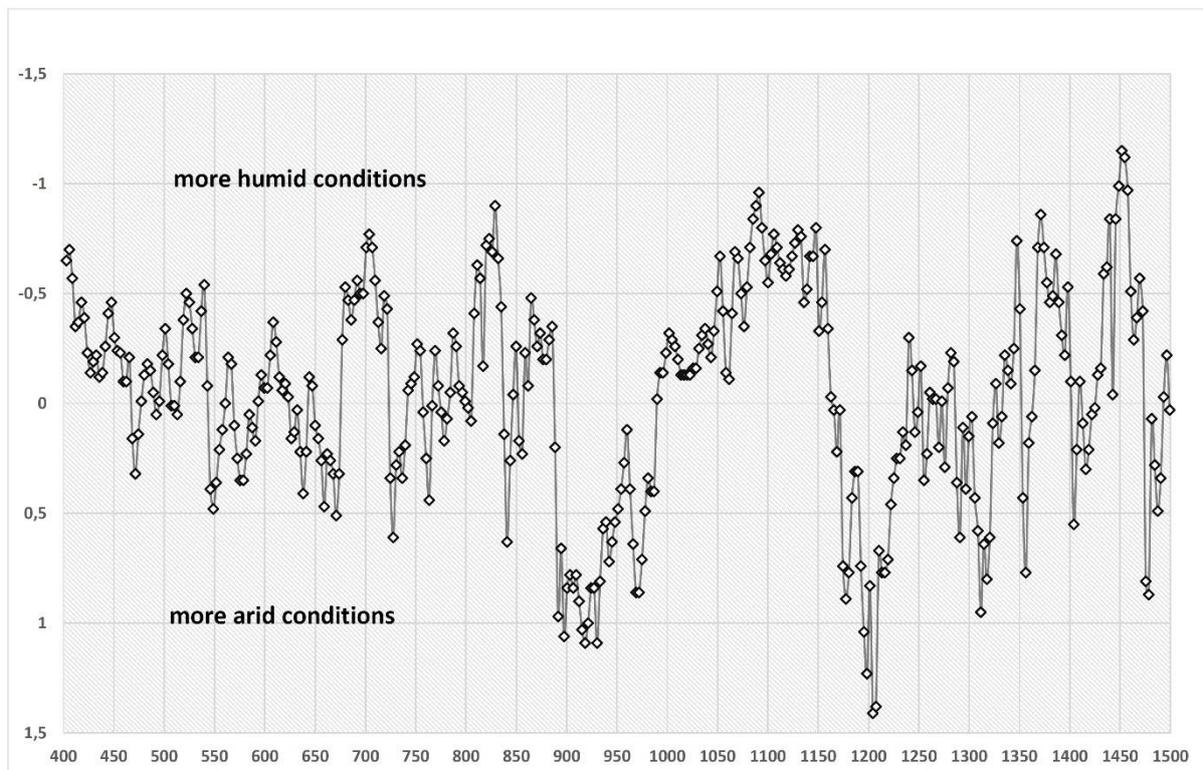

Fig. 6 Kuna Ba cave (Northern Iraq) oxygen isotopes record, 400–1500 CE (data: Sinha et al., "Role of climate"; graph: J. Preiser-Kapeller, OEAW, 2022)

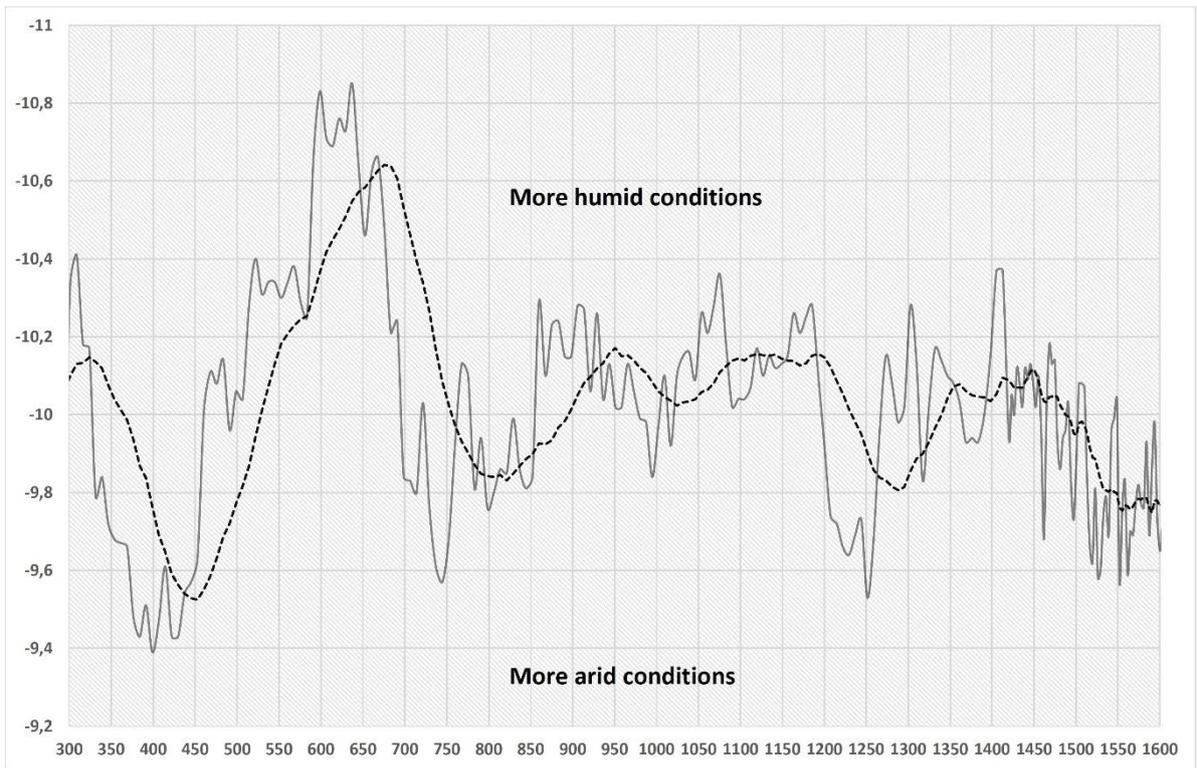

Fig. 7 Sofular Cave (northwestern Turkey) speleothem carbon isotopes record, 300–1600 CE; dotted line = moving average (data: Fleitmann et al., "Sofular Cave"; graph: J. Preiser-Kapeller, OEAW, 2022)

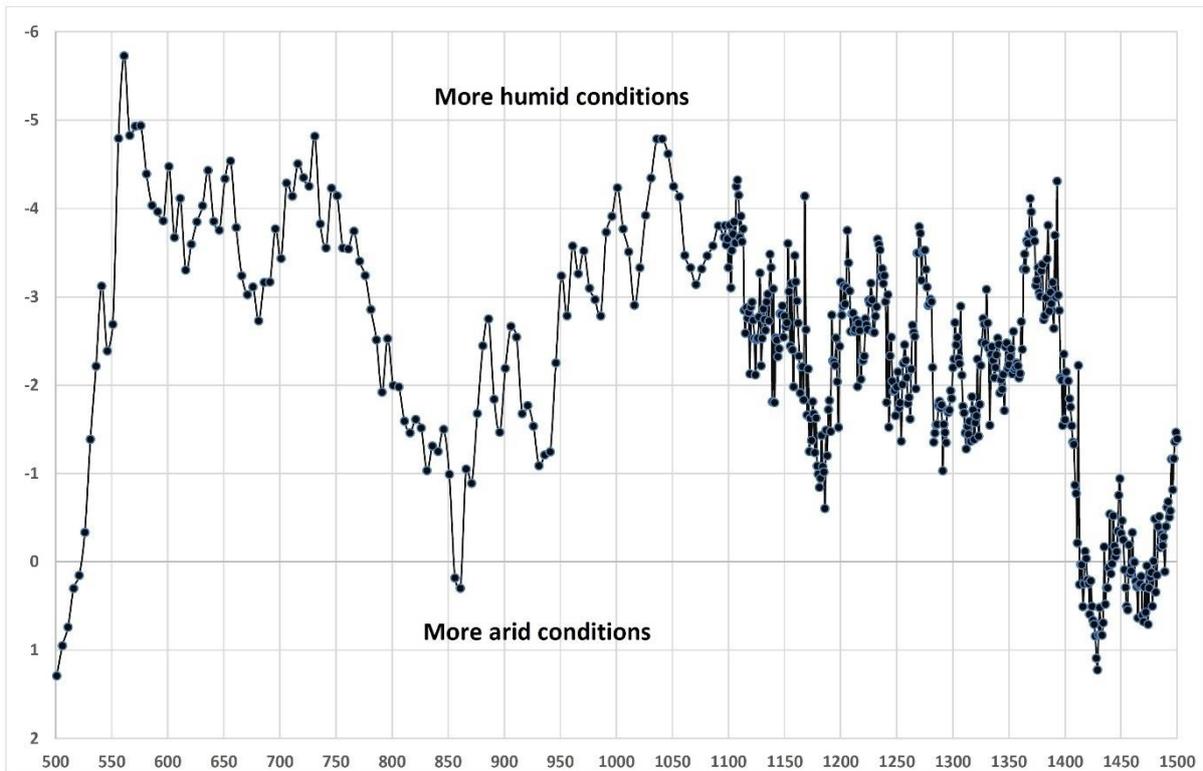

Fig. 8 Lake Nar (central Turkey) oxygen isotopes record, 500–1500 CE (data: Woodbridge/Roberts, "Late Holocene climate"; graph: J. Preiser-Kapeller, OEAW, 2022)

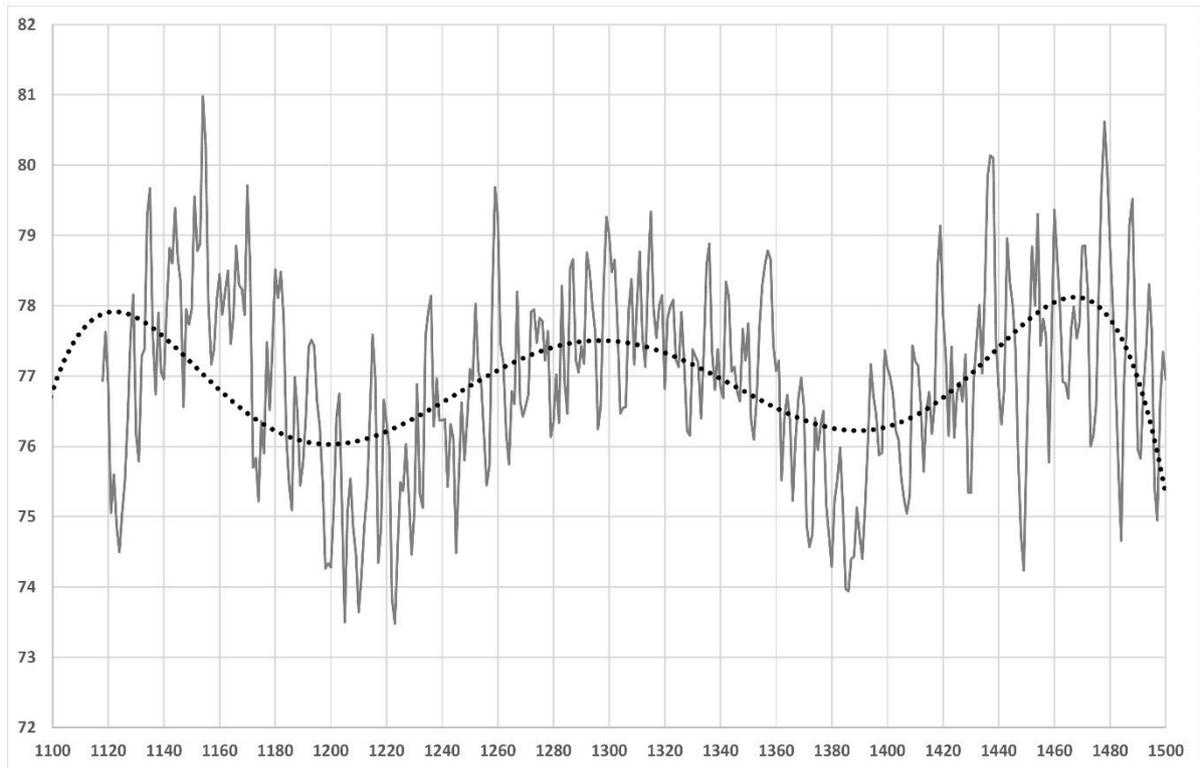

Fig. 9 Reconstructed May / June precipitation in mm for the northern Aegean region based on tree rings, 1100–1500 CE, 30-years average and long-term trend (data: Griggs et al., "A regional high-frequency reconstruction"; graph: J. Preiser-Kapeller, OEAW, 2022)